\documentclass[sigconf, screen]{acmart}
\PassOptionsToPackage{table,xcdraw}{xcolor}
\AtBeginDocument{%
  }

\copyrightyear{2025}
\acmYear{2025}
\setcopyright{acmlicensed}\acmConference[UIST '25]{The 38th Annual ACM Symposium on User Interface Software and Technology}{September 28-October 1, 2025}{Busan, Republic of Korea}
\acmBooktitle{The 38th Annual ACM Symposium on User Interface Software and Technology (UIST '25), September 28-October 1, 2025, Busan, Republic of Korea}
\acmDOI{10.1145/3746059.3747713}
\acmISBN{979-8-4007-2037-6/2025/09}
\acmPrice{}  % 去掉 $15.00

\newcommand{\tool}{\textsc{PaperBridge}~}
\newcommand{\toole}{\textsc{PaperBridge}}
\graphicspath{{figures/}{pictures/}{images/}{./}} % where to search for the images

\usepackage{amsmath}
\usepackage{algorithm}

\usepackage{color}
\usepackage{enumitem}
\usepackage{bbding}
\setlist{noitemsep,parsep=0pt,partopsep=0pt, leftmargin=10pt} 
\usepackage{listings}
\usepackage{multirow}
\usepackage{colortbl}
\usepackage{graphicx}
\usepackage{listings}
\newcommand{\rh}[1]{#1}

\newcommand{\eg}{{\it e.g.,\ }}

\newcommand{\ie}{{\it i.e.,\ }}
\definecolor{tablerowcolor}{rgb}{0.667,0.667,0.667 }
\definecolor{tablerowcolor2}{rgb}{0,0,0}
\definecolor{visual}{HTML}{e8efd9}
\definecolor{motion}{HTML}{fde7d5}
\definecolor{narrative}{HTML}{e2dce9}
\definecolor{audio}{HTML}{d6ebf2}
\definecolor{bluecrayola}{rgb}{0.12,0.46,1.0}

\begin{document}
\title[\toole: Crafting Research Narratives through Human-AI Co-Exploration]{\toole: Crafting Research Narratives through \\ Human-AI Co-Exploration}

% \title[]{PaperBridge: Inspiring Research Narrative Construction Through}

% \title[]{“I have over a dozen publications now, but...”: Inspiring Researchers to Curate Their Research Stories}

\author{Runhua Zhang}
\orcid{0000-0002-0519-5148}
\affiliation{%
  \institution{The Hong Kong University of Science and Technology}
  \city{Hong Kong}
  \country{China}  
}
\email{runhua.zhang@connect.ust.hk}

\author{Yang Ouyang}
\orcid{0009-0000-5841-7659}
\affiliation{%
  \institution{ShanghaiTech University}
  \city{Shanghai}
  \country{China}  
}
\email{ouyy@shanghaitech.edu.cn}

\author{Leixian Shen}
\orcid{0000-0003-1084-4912}
\affiliation{%
  \institution{The Hong Kong University of Science and Technology}
  \city{Hong Kong}
  \country{China}  
}
\email{lshenaj@connect.ust.hk}

\author{Yuying Tang}
\orcid{0009-0003-1906-2834}
\affiliation{%
  \institution{The Hong Kong University of Science and Technology}
  \city{Hong Kong}
  \country{China}  
}
\email{yuying.tang@connect.ust.hk}

\author{Xiaojuan Ma}
\orcid{0000-0002-9847-7784}
\affiliation{%
  \institution{The Hong Kong University of Science and Technology}
  \city{Hong Kong}
  \country{China}  
}
\email{mxj@cse.ust.hk}

\author{Huamin Qu}
\orcid{0000-0002-3344-9694}
\affiliation{%
  \institution{The Hong Kong University of Science and Technology}
  \city{Hong Kong}
  \country{China}  
}
\email{huamin@cse.ust.hk}

\author{Xian Xu}
\authornote {Corresponding author}
\orcid{0000-0002-2636-7498}
\affiliation{%
  \institution{Lingnan University}
  \city{Hong Kong}
  \country{China}}  
\affiliation{%
  \institution{The Hong Kong University of Science and Technology}
  \city{Hong Kong}
  \country{China}}
\email{xianxu0523@gmail.com}

% \orcid{1234-5678-7512}
% \affiliation{%

% \affiliation{%

\renewcommand{\shortauthors}{Zhang et al.}

\begin{abstract}

Researchers frequently need to synthesize their own publications into coherent narratives that demonstrate their scholarly contributions. To suit diverse communication contexts, exploring alternative ways to organize one’s work while maintaining coherence is particularly challenging, especially in interdisciplinary fields like HCI where individual researchers' publications may span diverse domains and methodologies. In this paper, we present \toole, a human–AI co-exploration system informed by a formative study and content analysis. \tool assists researchers in exploring diverse perspectives for organizing their publications into coherent narratives. At its core is a bi-directional analysis engine powered by large language models, supporting iterative exploration through both top-down user intent (\eg determining organization structure) and bottom-up refinement on narrative components (\eg thematic paper groupings). Our user study (N=12) demonstrated \toole's usability and effectiveness in facilitating the exploration of alternative research narratives. Our findings also provided empirical insights into how interactive systems can scaffold academic communication tasks. 

\end{abstract}




% \begin{figure*}[t]
%   \centering
%   \includegraphics[width=\linewidth]{figures/workflow.pdf}
%   \caption{Human-AI co-exploration on alternative research narratives with \toole.}
%   \label{fig:workflow}
% \end{figure*}

\begin{teaserfigure}
    \centering
    \includegraphics[width=0.98\linewidth]{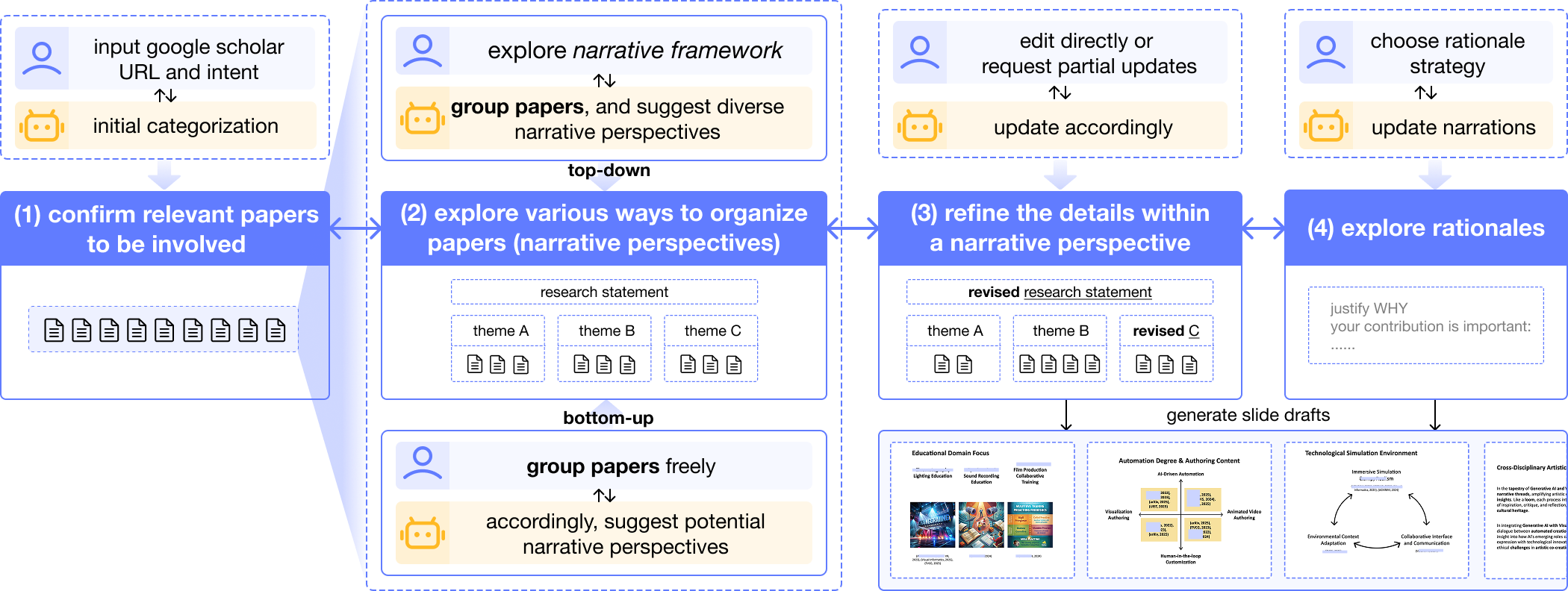}
    \caption{Human-AI co-exploration of alternative research narratives using \toole. (1) Users begin by selecting relevant papers they wish to organize into a coherent narrative. (2) \tool supports co-exploration of multiple narrative perspectives through both bottom-up and top-down approaches, allowing dynamic re-grouping of papers. (3) Once a promising narrative perspective is identified, users refine the details with \toole's assistance. (4) Finally, users examine rationale strategies to justify the significance of their perspective. In this workflow, users can move forward or backward flexibly, with slide drafts as the outputs of the exploration.}
\label{fig:workflow}
\end{teaserfigure}

\maketitle

\section{Introduction}

Academic communication plays a vital role in enabling researchers to disseminate findings, establish their scholarly identity, and cultivate collaborative relationships within their academic communities \cite{flowerdew2015identity, hyland2018narrative, aitken2010becoming, Heer, NarrativePlayer}. A key facet of this communication is the strategic organization of one's academic work—particularly publications—into coherent \textbf{research narratives} that clearly demonstrate a researcher’s contributions to the field \cite{hyland2011presentation, schulze2014finding}. Given the wide range of communication scenarios, from keynote speeches to dissertations and public outreach, researchers often need to craft \textbf{alternative narratives} tailored to specific audiences (such as academic departments, industry labs, or funding agencies) and communication goals, including job talks, grant proposals, and public engagement \cite{de2024delivering}. 

However, crafting alternative research narratives can be challenging. In addition to the need to tailor communication for different contexts, interdisciplinary fields like Human-Computer Interaction (HCI) add complexity, as research often spans diverse epistemologies, methodological traditions, and types of contributions \cite{thimbleby2004supporting}. For instance, even when a researcher's work centers on a single area such as creativity support, their projects may cut across multiple dimensions, from empirical studies to system development~\cite{Capel2023a}, while addressing various stages of creative practice (\eg ideation, implementation, iteration) across domains like video production, animation, design, and writing.

Moreover, unlike simply summarizing individual papers, constructing a research narrative requires identifying thematic structures across a body of work at an appropriate level of abstraction—one that is neither too broad nor too narrow \cite{cronon1992place, rosenthal2006narrated}. \rh{This can be particularly challenging for early-career researchers, who are just starting to shape their research trajectory. In addition,} as a researcher's publication record grows, this task increasingly demands reframing and reorganizing their work to form a coherent whole~\cite{Lee2015}. Despite these challenges, little attention has been dedicated to supporting researchers in synthesizing their own publications. This motivates our investigation into how to support narrative exploration, especially in the context of HCI. 

To address this gap, \rh{we conducted a formative study involving six early-career HCI researchers} to examine their current practices, key challenges, and expectations for potential support tools. We identified three essential elements in narrative construction: (1) \textbf{Narrative frameworks} that shape the overall flow or arc of the story, determining how subtopics are connected; (2) \textbf{Narrative perspectives} that provide specific lenses for highlighting contributions, typically consisting of a central contribution statement and associated thematic clusters that group related papers; and (3) \textbf{Narrative rationale} that explains and justifies why a particular perspective was chosen. Among these, the first two elements, which influence or are influenced by how papers are organized, posed the greatest challenges for participants. Based on these findings and their expectations, we derived four design considerations.

Before jumping into the design of \toole, we first conducted a content analysis of 53 publicly available HCI research talks to identify common practices that would help address our design considerations. This analysis revealed a design space with two key components: (1) Four common narrative frameworks: namely \textit{parallel}, \textit{linear}, \textit{circular}, and \textit{coordinate}. Each offers distinct logic for structuring relationships among publications, and (2) a set of storytelling strategies for the narrative rationale for justifying the significance of one's contribution.

Informed by design considerations from our formative study and the design space derived from content analysis, we present \toole, a human-AI co-exploration system that supports HCI researchers in exploring diverse perspectives to craft their research narratives. As shown in \autoref{fig:workflow}, the system implements a workflow that begins with researchers selecting the papers they wish to organize. Powered by a bi-directional analysis engine, \tool enables exploration of diverse narrative possibilities through both top-down and bottom-up approaches. In top-down exploration, researchers start by selecting a narrative framework, which guides the generation of multiple narrative perspectives—each composed of a contribution statement and thematic clusters with assigned papers. In bottom-up exploration, users can manually group papers and request matched perspectives based on their groupings. In both approaches, users can iteratively refine the system’s suggestions, either by editing directly or requesting partial updates, until they are satisfied with the resulting narratives. Finally, \tool also assists in articulating rationale to justify the chosen framing, and generates slide drafts that can be further revised or disseminated.

We evaluated \tool with 12 HCI researchers from various sub-domains. Participants responded positively to its usability and reported overall engaging experiences with \toole's co-exploration workflow. Our findings show that \tool not only improved the efficiency of exploring alternative research narratives, but also revealed how it scaffolded the co-exploration process, from initial trust establishment, to framework-based exploration, and engagement with keyword-driven perspective suggestions.

The contributions of this work are three-fold:
\begin{itemize}
    \item We conduct a formative study to gain a comprehensive understanding of how research narratives are constructed, leading to a design space encompassing four narrative frameworks and a set of rationale strategies.
    \item We present \toole, a human-AI co-exploration system that facilitates diverse research narrative exploration, which features a bi-directional analysis engine that enables both top-down and bottom-up approaches to narrative construction. 
    \item Through a user study with 12 HCI researchers, we demonstrate \toole's usability and reveal the co-exploration processes. We also discuss the implications for future academic communication support.
\end{itemize}

\section{Related Work}
In this section, we review prior work that supports academic profiling, research story construction for communication, and literature sensemaking. We articulate how our work builds on these efforts while identifying the gaps that motivate our research.

\subsection{Academic Communication}
Effective academic communication is fundamental to advancing research fields through knowledge dissemination and perspectives exchange \cite{flowerdew2015identity}. Researchers engage in this communication through various channels, including publishing papers, delivering conference talks, participating in academic workshops and seminars, and utilizing social media platforms \cite{de2024delivering}. Among these scenarios, presenting and profiling one’s research trajectory, narratives, or impact, plays a key role in building academic identity and establishing one's position within a research community \cite{aitken2010becoming, hyland2011presentation}.

To facilitate researchers' academic communication, prior research has explored constructing researcher profiles and stories by organizing personal and academic information, such as affiliations, education, and publication records \cite{yao2007unified, bravo2020researcher}. Various visualization systems have also been developed to represent researcher-related metrics, including publication volume, citation counts, collaboration networks, and topic evolution \cite{latif2018vis, 10.1145/2212776.2212796, wang2018visualizing, 7152908, kurosawa2011predicting}. For example, systems such as Citeology \cite{10.1145/2212776.2212796}, PivotPaths \cite{6327277}, PivotSlice \cite{6634163}, and PaperLens \cite{10.1145/1056808.1057069} enable multi-dimensional exploration of scholarly data, while CiteVis \cite{stasko2013citevis}, CiteRivers \cite{7192685}, and Wu et al.'s career trajectory visualization \cite{10.1145/2492517.2492638} provide overviews of research trends.

While these systems support researchers in promoting their academic impact through vivid, quantitative data-driven storytelling, they primarily focus on showcasing research breadth and external influence. On the one hand, this approach may limit researchers with fewer publications from effectively leveraging such tools. On the other hand, these systems offer limited support for helping researchers retrospectively reflect on their own publications to build research narratives grounded in the content and context of their work, which is a crucial process for demonstrating the depth of contribution within a particular field.

Although prior research has begun to explore content-level analysis of individual papers to support scholarly communication \cite{strobl2019digital}, few systems assist researchers in constructing a coherent and holistic reflection across their body of work. Yet, this process is critical in many academic scenarios, such as invited talks, job talks, seminars, and dissertations, where researchers may be expected to reorganize their work, identify meaningful connections, and tailor their narratives for different audiences and purposes \cite{schimel2012writing, hyland2011presentation}. Motivated by this gap, our work aims to explore how to support researchers in constructing their research narratives or profiles through the specific content and context of their own publications. 

\subsection{Literature Synthesis and Sensemaking}
Constructing research narratives based on one's own publications fundamentally relies on literature organization and information synthesis. While most prior research in HCI has focused on supporting literature sensemaking for external publications, \ie helping researchers explore, interpret, and synthesize knowledge scattered across others' papers, many of the underlying ideas and approaches also inspire our current work.

Broadly, prior studies on literature sensemaking can be categorized into two major threads. The first line of work focuses on fact-based information extraction and synthesis. Early systems often leveraged citation networks to help researchers identify relationships and patterns among a collection of papers \cite{choe2024fields, rachatasumrit2022citeread, he2019paperpoles}. For example, PaperQuest \cite{ponsard2016paperquest} utilized citation structures to assist literature review tasks. CiteSee \cite{chang2023citesee} augmented PDF reading experiences by highlighting citation relationships in context. In addition, tools such as Threddy \cite{kang2022threddy} and Synergi \cite{kang2023synergi} were designed to help users collect, organize, and understand research threads across multiple papers. Beyond citation structures, recent work also extracted factual content within papers, including tables, images, and textual claims, to enable comprehensive literature synthesis \cite{wang2024scidasynth}. Another line of work has begun to leverage large language models (LLMs) to synthesize paper content and help researchers contextualize the relationships among papers. For example, PaperWeaver \cite{lee2024paperweaver} generates synthesized descriptions to support paper recommendations, while DiscipLink \cite{zheng2024disciplink} enables interdisciplinary literature exploration by connecting concepts across fields.

Across these systems, a key insight is the combination of top-down and bottom-up workflows in supporting scholarly synthesis. Bottom-up approaches often help users incrementally grow their understanding of relevant literature by exploring specific papers, documents, or keywords \cite{learning2011apolo, ponsard2016paperquest, palani2023relatedly, kang2022threddy}. In contrast, top-down strategies aim to provide a high-level structural overview of a research landscape, enabling users to quickly grasp the organization of a domain \cite{shahaf2012metro}. A trend emerged and showed that the mixed-initiative approaches were explored \cite{kang2023synergi, zheng2024disciplink}. For example, Kang et al.'s  Synergi integrated both top-down and bottom-up approaches and further discusses how users' prior knowledge influences their preferences for top-down or bottom-up workflows \cite{kang2023synergi}.

Building on the insights from prior work, we adopt both top-down and bottom-up approaches to support researchers in constructing narratives from their own publications. On the one hand, researchers are often highly familiar with their own work, and therefore benefit from open-ended exploration that allows them to revisit and reflect on their ideas in flexible ways. On the other hand, the process of abstracting, reorganizing, and synthesizing familiar content into coherent research stories remains challenging. It may require top-down guidance or structural directions to help researchers identify meaningful patterns. Though existing systems for literature sensemaking have significantly advanced our understanding of information synthesis for text-heavy content \cite{overney2024sensemate}, they primarily focus on external literature exploration and comprehension. In contrast, constructing research narratives from one's own publications requires a different kind of sensemaking. Our work aimed at addressing this gap. 

\section{Formative Study}
We began by investigating how HCI researchers organize their research narratives, how they find alternatives, the challenges they face in current practices, and their expectations for a potential assistive tool, thereby identifying the design considerations.

\subsection{Methods}
\subsubsection{Participants}
Our formative study recruited six HCI researchers. Five participants were final-year PhD students (P1–P5), and one was a newly appointed postdoctoral fellow (P6). Their research backgrounds span diverse HCI subfields, including human–AI collaboration for creativity, data storytelling, visualization, virtual reality, HCI for science, and social computing. All participants had experience both attending and delivering academic talks at the time of the study.

\subsubsection{Procedures}
Each participant engaged in a semi-structured interview, including open-ended questions such as: ``\textit{Please describe your typical workflow to prepare your research narratives/stories delivered in academic occasions, such as giving talks}'', ``\textit{What are the considerations when preparing your research narratives?}'', and ``\textit{What challenges have you encountered when preparing such stories, and how did you address them?}'' Each session lasted around 40 minutes.

\subsubsection{Data Analysis}
We employed an open-coding approach \cite{holton2007coding} to inductively explore participants' current practices and challenges. Two authors independently reviewed the transcripts to generate initial codes, with subsequent discussions to resolve discrepancies and ensure coding consistency. Examples of the resulting codes include ``\textit{research narratives' components}'', ``\textit{paper organization methods}'', and ``\textit{storytelling intent}''. During the analysis, two professors with more than 15 and 20 years of experience in the HCI field, respectively, provided expert guidance and contributed to the establishment of analytical criteria that informed the organization and interpretation of the findings.

\subsection{Findings}
Our findings are organized into two main categories: (1) a hierarchical structure with three key elements for constructing research stories, and (2) key challenges identified in current practices, particularly associated with paper organization.

\subsubsection{Key Elements for Research Narratives and Corresponding Challenges}
%As shown in \autoref{fig:structure}, 
We identified three essential elements that HCI researchers consider when constructing research narratives: \textbf{narrative framework}, \textbf{narrative perspective}, and \textbf{narrative rationale}. Each of these elements plays a distinct role in how researchers organize and present their work. Our analysis also surfaced corresponding challenges that arise when researchers engage with these elements. Below, we describe each element and its associated challenge.

\textbf{(1) Narrative Framework} refers to the overarching structure or “flow” through which a researcher’s body of work is connected and presented. When crafting research narratives, participants often began by identifying central topics and reflecting on questions such as: “What is the main storyline across my work?” or “Can my papers be structured as addressing parallel challenges, or as a staged progression over time?” While some research trajectories lend themselves naturally to linear structures, many researchers actively experimented with alternative framings. For example, work on human-AI collaboration might be organized as a methodological journey (e.g., from identifying needs, to designing techniques, to system evaluation) or as an exploration of diverse contexts (e.g., creativity, decision-making, analytics).

\textbf{C1: Exploring Alternative Narrative Frameworks}. Participants shared that exploring different structural framings often meant rethinking how their papers related to one another. For example, P5 shared: ``\textit{Previously, I organized my papers into three clusters defined by different challenges. But then I saw others use staged progressions and tried that too. It didn’t quite work with my existing ideas.}'' Regarding this, the current approaches to discovering inspirations for relationships rely heavily on examining existing examples. P6 described searching through recorded talks to find models for research organization but noted that ``\textit{such a process was quite time-consuming and inefficient.}''

\textbf{(2) Narrative Perspective}. While frameworks outline a high-level structure, perspectives provide specific lenses or angles for spotlighting the researcher’s contributions. This statement typically synthesizes the fundamental research objective while highlighting the approach that might distinguish their work from others in the field. A typical perspective comprises: 

\begin{itemize}
    \item \textbf{A key contribution statement} (\eg ``My research advances human-AI collaboration by developing adaptive interaction techniques that support different collaborative scenarios''), which succinctly articulates the main takeaway or value proposition.
    \item \textbf{Thematic clusters grouping individual papers} (\eg ``Interaction techniques for creative tasks,'' ``Collaborative systems for decision-making,'' ``AI-assisted analytical workflows''). These clusters serve as concrete realizations of the contribution statement, each illustrating a particular dimension, approach, or facet of the overall claim.
\end{itemize}

\begin{figure}[t]
  \centering
  \includegraphics[width=0.8\linewidth]{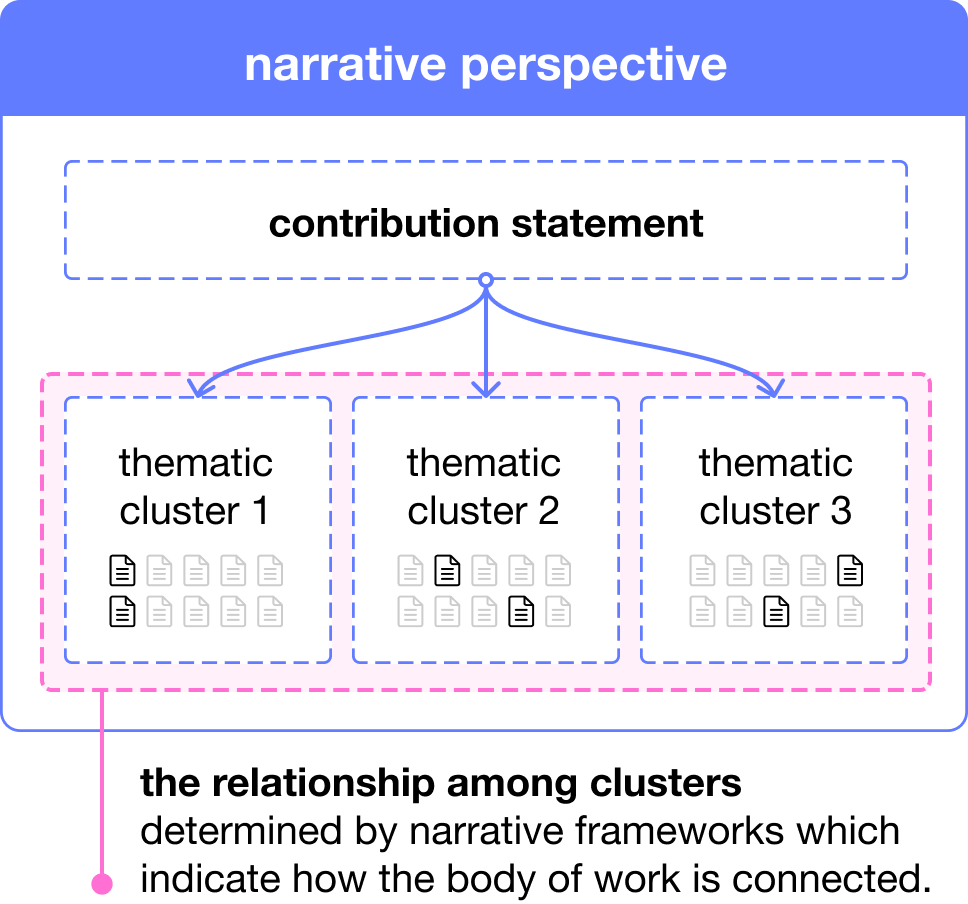}
  \caption{A narrative perspective is composed of three types of components: a contribution statement, thematic clusters, and individual papers, which are organized within the clusters to articulate the overarching themes.}
  \label{fig:perspective}
\end{figure}

The components of a narrative perspective and their relationships can be found in \autoref{fig:perspective}.

\textbf{C2: Exploring Alternative Narrative Perspectives.} Regarding narrative perspectives, participants consistently emphasized the interconnected challenges of articulating high-level contribution statements and organizing papers into meaningful thematic clusters. These reflect the core of research organization: contribution statements emerge from paper groupings, while effective groupings depend on clear contribution statements. Regarding contribution statements, participants struggled to capture unique intellectual contributions rather than merely describing research topics. P1, a final-year PhD student, explained: ``\textit{I know my research focus is to support creativity by supporting animation creation, and I am aware that each of my papers contributes to it. But I don't want to only emphasize my research is about creativity support, but how exactly my research contributes to it.}'' 

Simultaneously, participants struggled to develop thematic clusters that moved beyond surface-level categorizations to reveal deeper insights. P6 shared: ``\textit{There is a thread of my research regarding VR systems developed to understand different teaching and learning behaviors in such environments. Grouping my research by identifying the teaching content is simple, but I think it is somewhat too superficial to be insightful. However, I have not figured out how to group them from angles with more insightful concepts.}'' This difficulty becomes particularly acute when synthesizing work for comprehensive presentations, as P1 noted: ``\textit{To organize my thesis, I really want to abstract some more high-level concepts and form a line from my publications that will be presented in the thesis. Until now, I haven't figured it out even though I've talked a lot to my peers.}''

To address these interwoven challenges, participants reported seeking advice from peers or supervisors. However, these approaches were often considered inflexible, \rh{since they depended on others’ availability and were not well suited for iterative or self-paced narrative exploration}. Some attempted to use tools like large language models (such as ChatGPT) for assistance. \rh{Yet these general-purpose tools often required extensive manual refinement, and were commonly found to produce overly generic suggestions that lacked depth and contextual sensitivity, largely because they do not incorporate specific understanding of research narrative practices.}

\textbf{(3) Narrative Rationale}. The third element is the rationale, which provides justification for why the chosen perspective matters in the broader research context. It addresses questions like: ``\textit{Why is developing adaptive interaction techniques for human-AI collaboration important to HCI?}'' The rationale connects individual contributions to research challenges or social impact, helping researchers articulate not just what they did, but why their approach and findings are significant to advancing the field.

\textbf{C3: Adapting Narrative Rationale Accordingly}. Participants noted that when shifting between different perspectives, they needed to adjust their rationale to align with the new framing. As P4 explained: ``\textit{Depending on the scenario, I might emphasize the outputs of my visualization tools or the technical pipeline. Each choice requires me to rethink how I justify the significance of my contributions.}'' This challenge highlights the difficulty of not only selecting a narrative but also articulating its relevance in ways that resonate with different contexts (e.g., \textit{job talks}, \textit{theses}, \textit{research statements}).

\subsubsection{Expectations on An Assistive Tool} 
Our analysis also helped us identify three key expectations for an assistive tool that supports research narrative construction:

\textbf{Spark Inspirations, Not Complete Stories.} Participants expressed interest in tools that could provide inspirations for crafting narratives while preserving their agency in research interpretation. P4 described wanting ``\textit{automated suggestions for different organizational approaches, so that I can find different perspectives to understand my research.}'' This sentiment was balanced with a strong desire to maintain control over narrative development. As P2 explained: ``\textit{When discussing with peers or supervisors, they provide inspirations, and I take the decision on how to interpret their suggestions to form my research story, as I am the one who is most familiar with my research.}''

\textbf{Clear yet Flexible Workflow.} All participants characterized narrative crafting and paper organization as non-linear and often chaotic processes, expressing a desire for tools that could help streamline their workflow. P1 shared: ``\textit{I expected that the tool can help me with a clear workflow, so that I would not get lost in the ideas or thoughts.}'' However, participants emphasized that exploration should not come at the cost of flexibility. P4 noted: ``\textit{It should give me some freedom to refine its results and test my own ideas when organizing papers.}''

\textbf{Integration with Existing Workflows.} Beyond serving as a thinking aid, participants sought tools that could generate practical outputs compatible with their academic work. ``\textit{Being able to transform these narrative ideas directly into presentation slides, or at least some images would save significant time},'' noted P3. This reflected a desire for tools that could bridge the gap between narrative conceptualization and academic deliverables.

\subsection{Design Considerations}
\label{dcs}

Based on the elements, challenges, and expectations identified in our formative study, we propose four design considerations for supporting HCI researchers in exploring paper organization and crafting research narratives, with particular emphasis on facilitating explorations and inspiration to meet their expectations:

\textbf{DC1: Supporting Exploration of Diverse Narrative Frameworks.} To address C1, we aim to provide various narrative framework patterns that inspire researchers to explore different logical approaches to organizing their papers.

\textbf{DC2: Facilitating Alternative Narrative Perspective Discovery.} To address C2, we aim to support researchers in exploring diverse perspectives of their work. Besides, considering the expectations on flexible exploration, we aim to support them in iteratively refining the system’s suggestions or testing. 

\textbf{DC3: Supporting Narrative Rationale Development.}
As it is challenging to adapt the narrative rationale for different contexts (C3), we aim to assist researchers in articulating the significance of their chosen perspectives in different ways. 

\textbf{DC4: Generating Adaptable Visual Outputs.} To meet researchers' expectations for practical outputs, we aim to provide editable presentation slides that help researchers transform their narrative ideas into practical academic deliverables.

\section{Content Analysis}
As our formative study noticed that academic talks serve as a primary source of inspiration for research narratives, we conducted a content analysis of public HCI guest talks to address our design considerations, especially DC1 and DC3. Through this analysis, we examined (1) different approaches to organizing papers into narrative frameworks and (2) rationale strategies for justifying the significance of contribution statements.

\begin{figure*}[t]
  \centering
  \includegraphics[width=\linewidth]{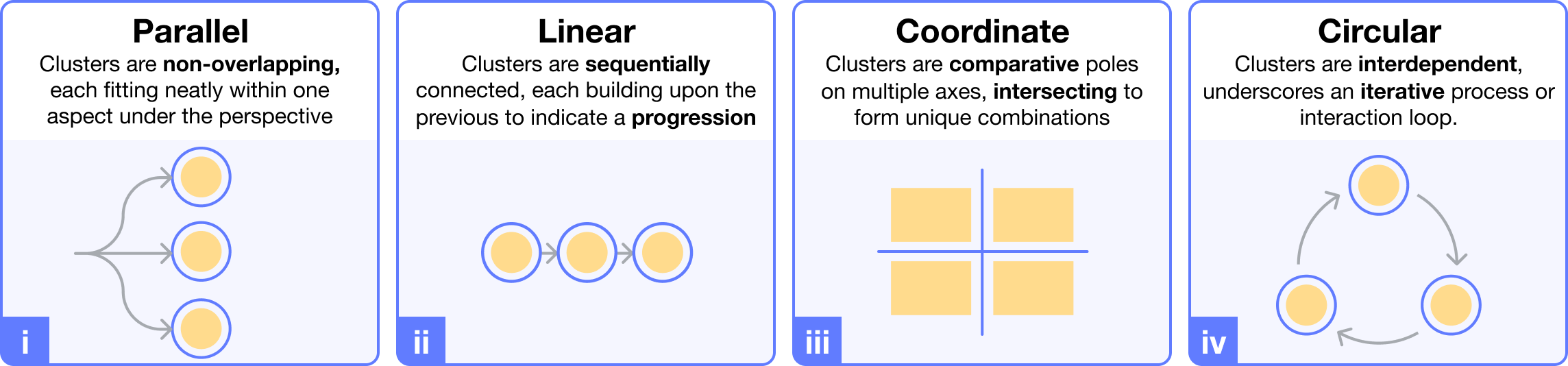}
  \caption{Four common narrative frameworks identified from our content analysis. Each framework describes a distinct way of organizing and connecting a researcher's body of work around a central storyline.}
  \label{fig:patterns}
\end{figure*}

\subsection{Methods} 
\subsubsection{Dataset}
We assembled a dataset of 53 publicly accessible job talks and guest seminars \footnote{\rh{See  \url{https://docs.google.com/spreadsheets/d/1DfnBhTOpjPVJyibi6dTsiTnCkhj8zSZl/edit?usp=sharing&ouid=102936865702623175143&rtpof=true&sd=true} for 53 public HCI talks used in our analysis}} hosted by prominent HCI communities, such as the CMU HCII Seminar Series, the UW CS Seminar Series, Stanford Online, etc. These sources invite researchers from various cultural and research backgrounds, ensuring that our corpus captures a broad spectrum of HCI research domains. Our inclusion criteria were: (1) The speaker had an academic affiliation rather than an industry role; (2) The talk’s content drew upon a series of publications, and (3) The talk video included visual presentation materials (\eg slides).

\subsubsection{\rh{Preliminary Explorations}}
We began with a pilot analysis of 10 talk videos to refine our focus. This initial phase helped us identify the part where researchers build the “big picture” of their research stories. This part usually appeared in the first 10–15 minutes of the talk and included presenting the research topics and motivation, stating the contribution perspective (or proposing key research questions or gaps), and showing how their publications supported these ideas. We found this part more relevant to our study than the detailed discussion of individual projects. Therefore, we decided not to analyze the project sections in detail, as our main goal was to understand how researchers construct the ``big picture'' of their research story.

\rh{In addition, we observed diverse approaches to constructing narrative frameworks. According to insights from our formative study, the key of such frameworks lies in how researchers articulate the connections among their publications to collectively address a shared research problem. Building on this understanding, our observations identified recurring framework patterns, such as illustrating how individual works tackled distinct facets of a broader agenda, or how later publications built upon earlier ones. Based on these patterns, our analysis further examined the underlying relationships among publications as framed by the researchers.}

\subsubsection{\rh{Procedures}} Following the pilot, two authors independently coded the talk videos using an inductive thematic analysis approach, focusing on the two focal points outlined earlier \rh{and the preliminary insights collected}. The disagreements were resolved through rounds of discussion. Finally, two senior HCI researchers were invited to be involved in the discussion to refine the patterns and strategies that had emerged.  

\subsection{Findings}
\subsubsection{Four Patterns for Narrative Frameworks} 
\label{frameworks}
\rh{Our findings first addressed DC1 by identifying the common narrative frameworks. As shown in \autoref{fig:patterns}, we identified four patterns used to organize and connect publications: \textit{parallel}, \textit{linear}, \textit{coordinate}, and \textit{circular}. Each pattern shaped a distinct narrative arc within research trajectories. The visual representations of these patterns were informed by the tree and matrix diagrams that researchers employed.} Specifically, the publications were organized into different thematic clusters, and the different relationships among these clusters define each pattern. Below, we define each of them, describe its characteristics, and provide examples.

\textbf{Parallel Structure: Multiple Facets of Contribution.}
The parallel structure (\autoref{fig:patterns}-i) represents the most common approach (43 out of 53) used by HCI researchers to organize their papers. In this structure, thematic clusters show a non-overlapping relationship and represent distinct technical or methodological approaches or different perspectives for formulating a problem space that collectively addresses a central research challenge. For example, a researcher investigating \textit{accessible design} might organize papers into clusters focused on \textit{vision-based accessibility}, \textit{audio accessibility}, and \textit{haptic feedback}. Each represents a different sensory modality but together serving the main topic.

\textbf{Linear Structure: Progressive Development.}
Linear structures (\autoref{fig:patterns}-ii) appeared five times in our clusters (5/53). In this structure, the thematic clusters illustrate a sequential flow where papers are organized in a progressive line, highlighting how each stage builds upon insights from previous stages. For instance, a researcher studying \textit{embodied interaction} might organize papers into clusters showing progression from \textit{sensing technologies} to \textit{interaction techniques} to \textit{application implementations}. The linear structure is ideal for showcasing how initial explorations led to sophisticated outcomes or how works span different levels of advancement.

\textbf{Coordinate Structure: Comparative Dimensions.}
In coordinate structures (4 out of 53, \autoref{fig:patterns}-iii), thematic clusters are positioned within a conceptual space defined by key dimensions that illustrate fundamental tensions in the HCI community, such as ``\textit{user control vs. automation}'' or ``\textit{expressiveness vs. efficiency}.'' This structure operationalizes what HCI researchers often refer to as a design space. The coordinate structure helps research statements emphasize the exploration of design spaces or tensions between competing factors, providing a systematic way to present how works exist at the intersections between concepts.

\textbf{Circular Structure: Iterative Refinement.}
In circular structures (1 out of 53, \autoref{fig:patterns}-iv), each thematic cluster represents an interconnected phase that continuously informs others, serving the research statement through iterative processes or feedback loops. The circular structure is particularly valuable for research grounded in iterative processes, design thinking methodologies, or user-centered development cycles. It effectively communicates how different facets of research mutually reinforce each other and how complex problems require recursive refinement.

\subsubsection{Rationale Strategies}
\label{storytelling}
In addition to the four common narrative frameworks, we also identified a comprehensive set of rationale strategies for justifying why a researcher's contribution statement is important to a domain. 

As presented in \autoref{tab:what_why_strategies}, the strategies were categorized according to Aristotle's rhetorical framework \cite{aristotle1954rhetoric} of \textbf{ethos}, \textbf{pathos}, and \textbf{logos}, which researchers strategically deploy to justify their work. \textbf{Ethos} refers to building credibility and trust — convincing audiences that the authors are knowledgeable and their work is grounded in recognized expertise. For establishing ethos, researchers utilized literature endorsement (8 instances), highlighted industry/academic attention (2), and incorporated quotations from prominent figures in the field (6). \textbf{Pathos} appeals to emotion and shared values — helping audiences connect with the problem on a personal or societal level. To engage pathos, researchers shared personal experiences (3), referenced relevant social events (5), aligned with common values and ethical considerations (7), and emphasized public demand or popularity (1). \textbf{Logos} relies on logical reasoning and evidence — using facts, data, and analysis to demonstrate the importance of the problem. Logos-based strategies included presenting data with visualizations (7), identifying historical patterns and trends (8), demonstrating tangible impact (9), and articulating the consequences of inaction (1). It is worth noting that researchers frequently employed multiple strategies simultaneously to justify their contribution, resulting in a total count of more than 53. The detailed definitions and examples of each strategy are provided in \autoref{tab:what_why_strategies}. This collection of storytelling strategies informs our approach to addressing DC3.

\begin{table*}[t]
\centering
\renewcommand{\arraystretch}{1.2}
\setlength{\tabcolsep}{6pt}
\begin{tabular}{p{4cm}p{13cm}}
\toprule
\textbf{Strategy} & \textbf{Method and Definition} \\  
\midrule
\textbf{Ethos} (establish trustworthiness and expertise) 
    & \textbf{Literature Endorsement}: Cite established research that underscores the importance of this gap. \newline
      \textbf{Industry/Academic Attention}: Show how academia or industry prioritizes this topic through funding, discussions, or publications. \newline
      \textbf{Big Name's Quote}: Use expert endorsements to reinforce the significance of the problem. \\
\midrule
\textbf{Pathos} (appeal to emotions and values) 
    & \textbf{Personal Experience}: Share a real-life story to make the issue more relatable and engaging. \newline
      \textbf{Social Events}: Link the research to societal, political, or cultural events to highlight its relevance. \newline
      \textbf{Common Values \& Ethics}: Appeal to widely shared moral or ethical principles to justify the importance of this research. \newline
      \textbf{Public Demand \& Popularity}: Highlight growing public interest or widespread adoption to demonstrate the topic's timeliness. \\
\midrule
\textbf{Logos} (appeal to logic and evidence) 
    & \textbf{Data \& Visualization}: Use statistics or visual evidence to illustrate the scale or urgency of the problem. \newline
      \textbf{Historical Patterns \& Trends}: Show how this issue fits within broader technological advancements or ongoing trends. \newline
      \textbf{Demonstrating Impact}: Emphasize the tangible, long-term benefits of addressing this problem. \newline
      \textbf{Consequences of Inaction}: Highlight risks, negative outcomes, or missed opportunities if the problem is ignored. \\
\bottomrule
\end{tabular}
\caption{Rationale strategies used by researchers to justify the importance of their contributions. These are categorized following Aristotle’s rhetorical modes: \textit{ethos} (credibility), \textit{pathos} (emotional appeal), and \textit{logos} (logical reasoning).}
\label{tab:what_why_strategies}
\end{table*}

\section{\toole}

Based on the insights from our formative study and content analysis, we present \toole, a human-AI co-exploration system to inspire HCI researchers to craft their research narratives by supporting them in exploring diverse perspectives to organize their publications. Specifically, we first introduce \toole's design and the corresponding design considerations through a system overview. Then, we present a detailed usage scenario to walk through how researchers can interact with \toole. Finally, we describe the backbone of \toole, a bi-directional analysis engine that enables both top-down and bottom-up explorations. 

\begin{figure*}[t]
  \centering
  \includegraphics[width=\linewidth]{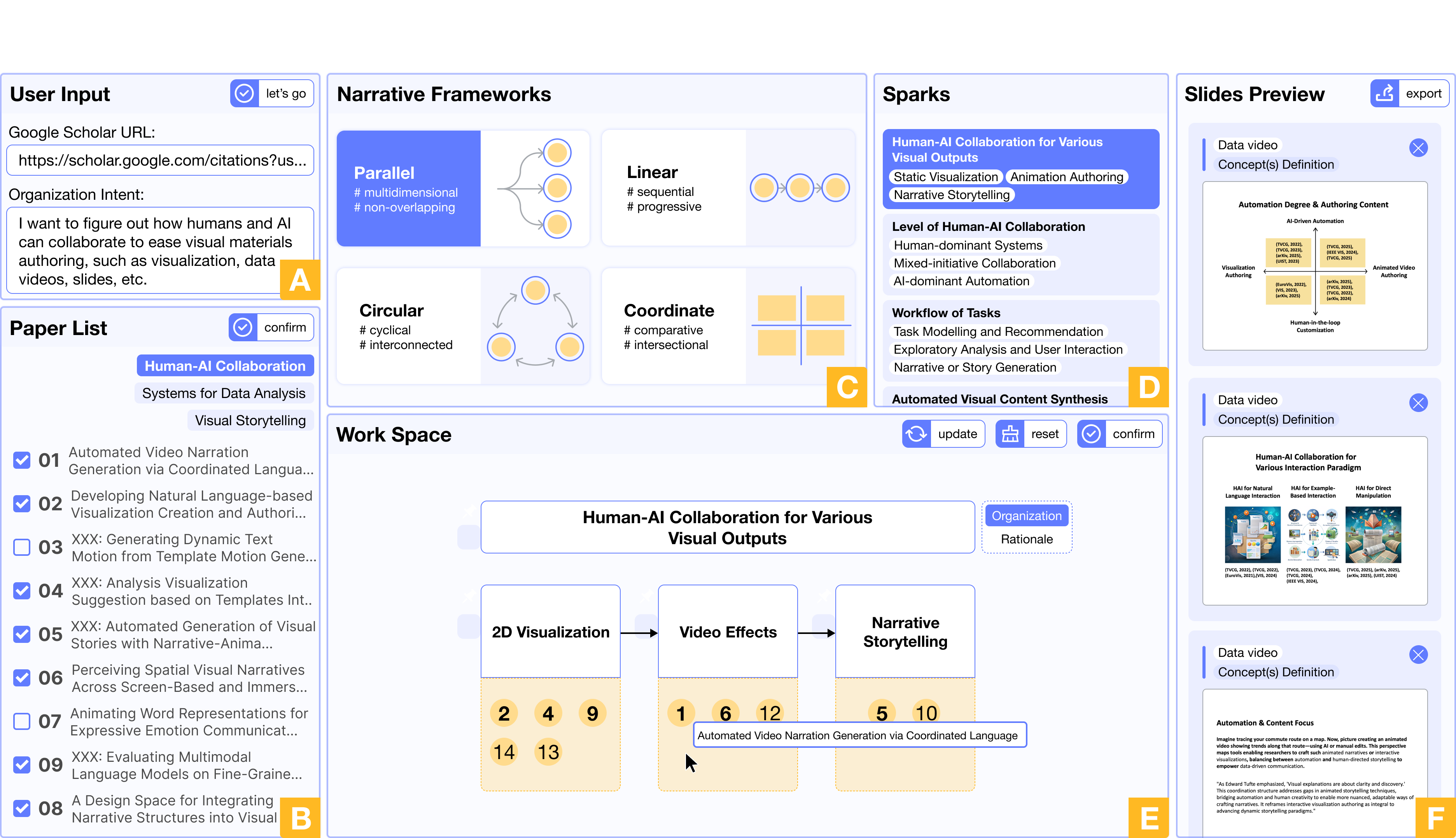}
  \caption{\tool can be navigated through left, middle, and right panels. It supports (1) Paper Management in Panels A \& B on the left, (2) Narrative Exploration (main feature) in Panels C, D, and E in the center, and (3) Slide Draft Preview in Panel F on the right.}
  \label{fig:main}
\end{figure*}

\subsection{Overview}

As shown in \autoref{fig:main}, \tool consists of three main components that collectively address four key design considerations (DCs): Paper Management (Panels A \& B, left view), Narrative Exploration (Panels C, D, and E, center view), and Slide Draft Preview (Panel F, right view).

The workflow begins with paper \textbf{selection stage} in Panels A and B (\autoref{fig:main}). After entering the user's Google Scholar URL and specifying their narrative intent, \tool automatically retrieves their publications and generates initial, topic-based categorizations. This allows researchers to browse their publication corpus efficiently and iteratively refine both the selected papers and the category labels to align with their intended narrative scope.

Once users have chosen the papers they wish to include, they move to the Narrative Exploration area in the center (Panels C, D, and E, \autoref{fig:main}).

\textbf{Narrative Framework Exploration in Panel C (DC1)}. This panel presents four common narrative frameworks (\textit{parallel}, \textit{linear}, \textit{circular}, and \textit{coordinate}) identified in \autoref{frameworks}. By switching among these frameworks, users can explore different high-level structures to see how their selected publications interconnect. 

\textbf{Narrative Perspectives Exploration in Panel D (DC2)}. Selecting a framework triggers \toole’s analysis engine to produce “sparks”: potential narrative perspectives that fit the chosen framework. Each spark includes a candidate contribution statement and one or more thematic clusters, giving users alternate angles for framing their research story. The sparks are designed to be presented in a keyword format, encouraging users to interpret the narratives independently first.

\textbf{Perspectives Refinement and Rationale Exploration in Panel E (DC2 \& DC3)}. Panel E is a workspace that provides two modes that can be switched with the button alongside the contribution statement text box. 
\begin{itemize}
    \item One is \textbf{Organization Mode} (Panel E, \autoref{fig:main}), where users can check and refine the components in each spark (\ie narrative perspective) suggested by \tool through a top-down analysis. Bottom-up exploration also happens in Panel E, where users assign papers to clusters manually, then have \tool synthesize contribution statements and cluster themes. In either way, users can refine the components, such as editing statements and clusters' themes, and dragging and dropping papers into different clusters. They can also confirm elements they like and request the regeneration of anything unsatisfactory. To facilitate the exploration, papers are represented as numbered circles with a hover-activated title display, and contextual explanations on the elements are accessible through hover interactions.
    \item Another is \textbf{Rationale Mode} (see \autoref{fig:rationale} where this mode is activated), where users can explore various rationale strategies (identified in \autoref{storytelling} and displayed in Panel C) by dragging and dropping them into Panel E, helping them articulate the significance of their chosen perspective.
\end{itemize}

Throughout the workflow, users can switch freely among frameworks (Panel C), perspectives (Panel D), workspace (Panel E) to explore diverse narratives that suit their objectives. Whenever satisfied with the content in work space, they can confirm their workspace content, and \tool will transform the content into editable slides \textbf{(DC4)} and provide a preview in Panel F (\autoref{fig:main}). Users can export them for further refinement.

\subsection{Usage Scenarios Walkthrough}

In this section, we present a usage scenario that walks through \toole, illustrating how it facilitates co-exploration on research narrative based on user intent and interactions. Below, we use \textit{italics} to indicate \textit{Flora's thoughts}, and \texttt{teletype} to indicate the \texttt{button} on the interface. We use ``Initial Capitalization'' for the content suggested by \tool to the user. 

\textbf{Scenario Background.}
Imagine Flora\footnote{To comply with anonymity requirements during review, we constructed this scenario using a synthetic profile. We combined publications from multiple researchers, reframed paper titles, altered venues and publication years, and removed identifying system names. The context here and the content shown in \autoref{fig:main} do not reflect any single researcher's actual publication history.}, a postdoctoral fellow in HCI who focuses on visual storytelling, with an emphasis on motion effects. 
Her recent research portfolio has expanded significantly with new publications related to human–AI collaboration (\eg the underlying collaboration paradigms to achieve the outputs). She wants to reorganize these works under this broader topic but struggles to pinpoint suitable perspectives. Flora turns to \tool for inspiration. 

\textbf{Paper Selection (Panels A \& B).}
As shown in \autoref{fig:main}-A, after entering her Google Scholar URL into \toole, Flora types her current intent:
\textit{``I want to figure out how humans and AI can collaborate to ease visual materials authoring, such as visualization, data videos, slides, etc.''} 
She clicks the \texttt{let’s go} button. In response, \tool automatically extracts her publications from Google Scholar and categorizes them into three clusters (\autoref{fig:main}-B): ``Human–AI Collaboration''; ``Visual Storytelling''; ``Systems for Data Analysis''. \textit{“These categories look pretty close to my mental map of my work,”} Flora notes as she navigates the tags in Panel B. She checks the relevant papers potentially relevant to Human-AI collaboration, and refines ``Human-AI Collaboration'' to ``Human–AI Collaboration for Visual Storytelling'' to better align with her intended focus. Satisfied with these selected papers, she clicks \texttt{Confirm} in Panel B.

\begin{figure*}[t]
  \centering
  \includegraphics[width=\linewidth]{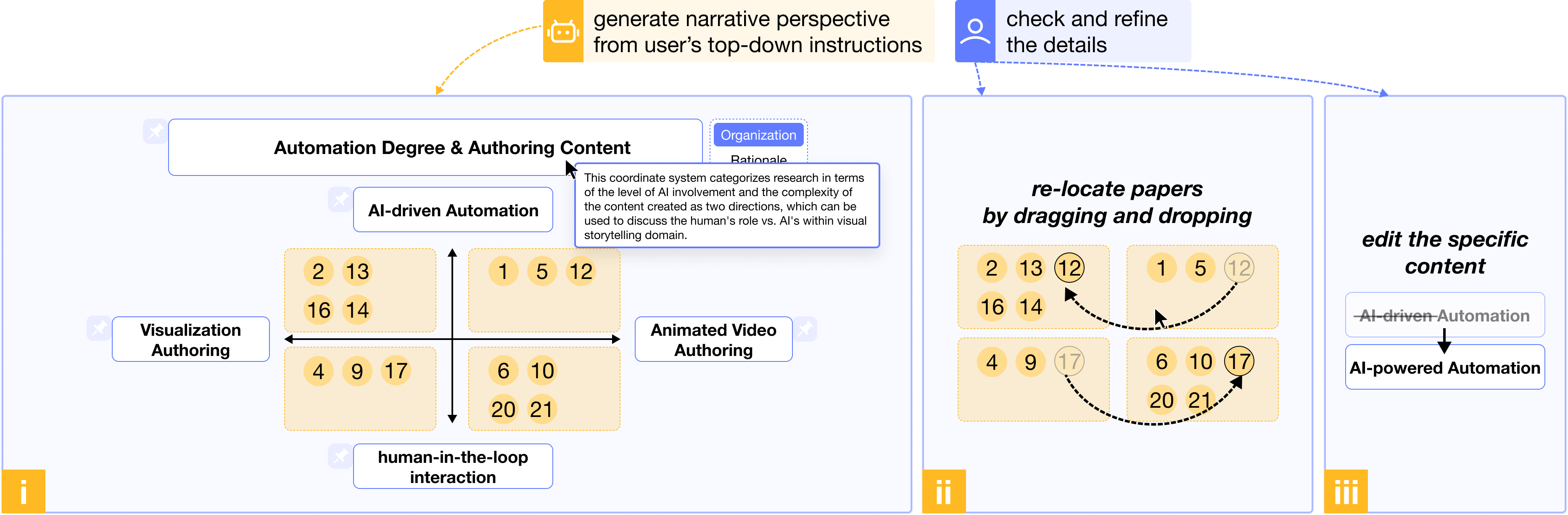}
  \caption{\tool supports user in checking, adjusting, and revising the top-down exploration results. (i) One narrative perspective suggested by \toole, including contribution statement, thematic themes, and assigned papers. The user can hover on the components to see the explanations from \toole. (ii) User can re-locate papers to other clusters by dragging and dropping interaction. (iii) The user can edit the specific content to satisfy their intent.}
  \label{fig:corrdinate}
\end{figure*}

\textbf{Exploring Potential Narrative Perspectives (Panel C, D, \& E).}
Flora then moved to the middle area. She starts in Panel~C (\autoref{fig:main}-C), which displays four different narrative frameworks: \texttt{parallel}, \texttt{linear}, \texttt{circular}, and \texttt{coordinate}. \textit{``My prior approach seems like a linear approach,''} Flora reflects. She clicks the \texttt{Linear} first to see how \tool maps her publications in a linear arc. \tool responds by generating ``sparks'' in Panel D (\autoref{fig:main}-D). The first spark closely matches her previous storyline, though phrased differently: ``Human–AI Collaboration for Various Visual Outputs,'' with thematic clusters labeled ``2D Visualization,'' ``Video Effects,'' and ``Narrative Storytelling''. She sees that hovering over the numbered circles in Panel E reveals each paper’s title (\autoref{fig:main}-E). \textit{``Good. These align with the developmental path of my work—starting with 2D data viz and moving into animations,''} Flora thinks. Encouraged by this familiarity, she decides to explore more potential perspectives on human–AI collaboration with the help of \toole.

Flora clicks the \texttt{Coordinate} framework next, and \tool analyzes her publications again. The first candidate perspective in Panel D is labeled ``Automation Degree \& Authoring Content''. As shown in \autoref{fig:corrdinate}-i, this perspective is mapped onto a coordinate plane with two dimensions: ``AI-driven Automation'' vs. ``Human-in-the-loop Interaction'' on one axis, and ``Visualization Authoring'' vs. ``Animated Video Authoring'' on the other. \textit{``This is interesting,''} Flora thinks, \textit{``It distinguishes how AI’s role varies with different output formats.''} She hovers the perspective text box to reveal \toole’s explanation (\autoref{fig:corrdinate}-i): \textit{``This coordinate system categorizes research in terms of the level of AI involvement and the complexity of the content created as two directions, which can be used to discuss the human's role vs. AI's within visual storytelling domain.''} \textit{“That’s spot on!”} Flora thinks. She checks the assigned papers by hovering over their numbered circles and relocates two papers by dragging and dropping them into another cluster (\autoref{fig:corrdinate}-ii). Besides, she revises ``AI-driven Automation'' to ``AI-powered Automation''(\autoref{fig:corrdinate}-iii). With that small tweak done, she clicks \texttt{Confirm}, and \tool generates a corresponding slide reflecting this coordinate-based perspective (first slide shown in Panel F, \autoref{fig:main}).

\begin{figure*}[t]
  \centering
  \includegraphics[width=\linewidth]{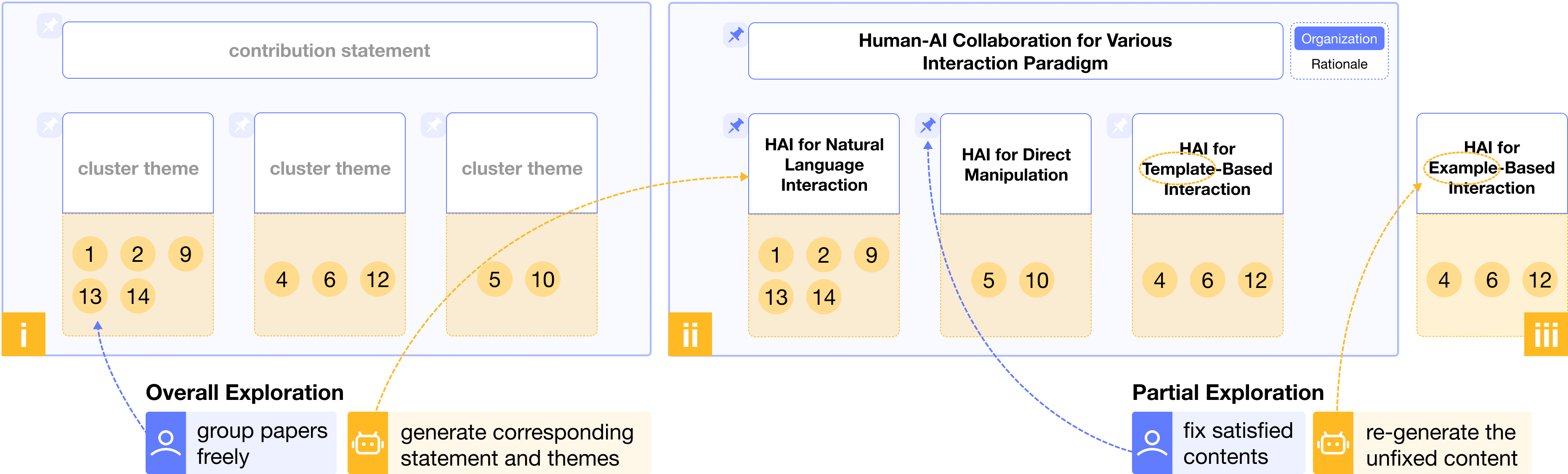}
  \caption{\tool supports bottom-up explorations in both overall or partial manner. (i) User can group their papers freely, and request \tool to generate overall suggestions. (ii) \tool suggests the corresponding narrative perspectives, including contribution statement and clusters' themes. (iii) User can fix the satisfactory content and request \tool to re-generate the unfixed content.}
  \label{fig:parallel}
\end{figure*}

\textit{``Maybe there are other angles worth trying,''} Flora thinks. She decides to explore a bottom-up approach under the \texttt{Parallel} framework. This time, she first clicks \texttt{reset} button in Panel E, and manually groups her papers based on her understanding of the common and distinct traits in her research (\autoref{fig:parallel}-i). She then requests \tool to generate suitable contribution statements and cluster themes by clicking the \texttt{update} button in Panel E. Among the various attempts, one especially catches her eye (\autoref{fig:parallel}-ii): ``Human–AI Collaboration for Various Interaction Paradigms,'' yielding three thematic clusters: ``HAI for Natural Language Interaction'', ``HAI for Template-based Interaction'', and ``HAI for Direct Manipulation Interaction''. \textit{“I never thought of it this way, but it nicely highlights how I’ve been supporting diverse interaction methods,”} she realizes. She re-checks the paper assignments in each cluster, adjusts a few accordingly, and decides ``Template-based Interaction'' feels a bit vague. She locks the other two themes as final and clicks \texttt{update} in Panel E again. \tool regenerates that theme as ``HAI for Example-based Interaction'' (\autoref{fig:parallel}-iii), which Flora appreciates. She confirms it, prompting \tool to generate another slide (second slide shown in Panel F, \autoref{fig:main}).

\begin{figure*}[t]
  \centering
  \includegraphics[width=\linewidth]{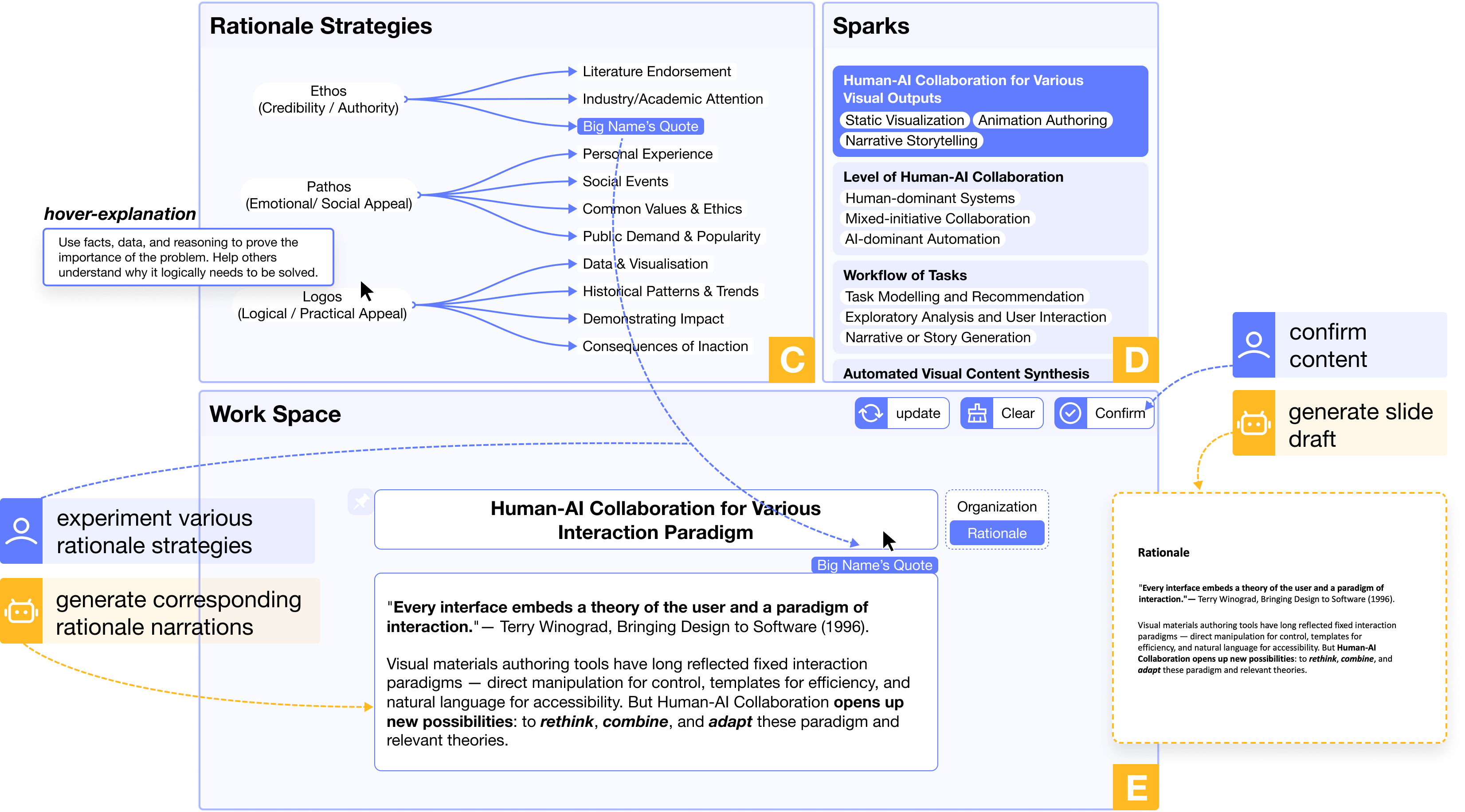}
  \caption{\tool supports users in exploring various rationale strategies to justify the significance of the specific contribution statement by generating corresponding narration drafts for users. }
  \label{fig:rationale}
\end{figure*}

\textbf{Crafting the Rationale (Panel E).} Though Flora is satisfied with the interaction paradigm in organizing her research, she is not sure how to justify its importance in a talk. Therefore, Flora clicks the \texttt{Rationale} button in Panel E, enabling Rationale Mode. As shown in \autoref{fig:rationale}, simultaneously, \tool displays various rationale strategies in Panel C. \textit{``Let's see what with big name's quote,''} she thinks. She drags the \texttt{big name's quote} strategy to the text boxes in Panel E. \tool automatically generates the narrations based on this strategy and shows the content in the text box (see \autoref{fig:rationale}). Clicking \texttt{Confirm}, she sees the current content in the workspace is organized into a new rationale slide. 

\textbf{Exporting the Slides (Panel F).}
Finally, Flora compares the different frameworks and saves several promising perspectives. \textit{``I’ve got a nice variety of narratives,''} she thinks. When she feels she has gathered enough directions, she clicks the \texttt{Export} button in Panel~F (\autoref{fig:main}-F). \tool generates a slide file with scripts (in \textit{.pptx} format). \textit{``Perfect. I can show these different ways of framing my work to my advisor,''} Flora thinks, \textit{``This might spark new ideas and feedback.''}

In this way, Flora capitalizes on both top-down and bottom-up exploration, stepping freely between different frameworks, customizing thematic clusters, and refining her rationale. \toole’s iterative workflow helps her craft diverse narrative perspectives that reflect her research contributions in new and insightful ways.

\begin{figure*}[t]
  \centering
  \includegraphics[width=\linewidth]{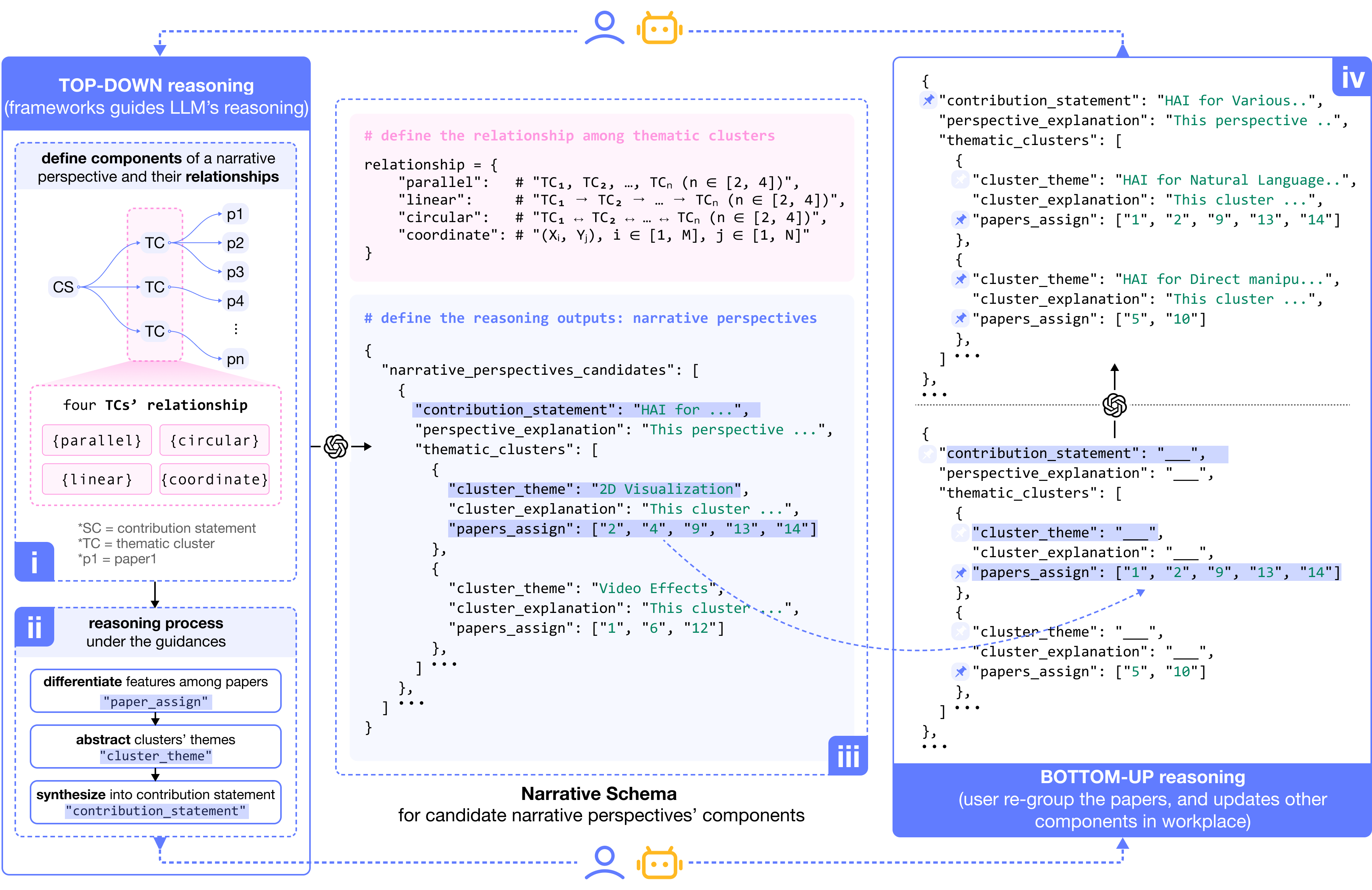}
  \caption{Backend implementation for \toole, a bi-directional analysis engine that supports top-down and bottom-up reasoning for organizing and synthesizing publications (abstracts and titles) into narrative components. \textit{Top-down}, we define structured narrative components, especially inter-cluster relationships, to guide LLMs in grouping thematically coherent paper clusters. \textit{Bottom-up}, users can re-group papers themselves, refine clusters, and reshape contribution perspectives. The \textit{Narrative Schema} serves as a bridge between both reasoning processes by storing and updating narrative components.}
  \label{fig:backend}
\end{figure*}

\subsection{Backend Engine: Bi-directional Analysis}

The backend of \tool is designed to support users in constructing research narratives from their collection of publications. It consists of four modules: (1) paper collection and sifting, (2) bi-directional analysis engine as the core, (3) evaluation and ranking, and (4) slide generation. We describe each module in detail below.

\subsubsection{Paper Collection and Sifting} 
Given a Google Scholar URL provided by the user, we use the scholarly package\footnote{See \url{https://pypi.org/project/scholarly/} for documentation.} to retrieve the author’s publication metadata, including each paper’s title, abstract, citation string, and unique ID. To support efficient navigation and selection, the system leverages LLMs (OpenAI's o3-mini) to categorize the retrieved papers by topic based on user intent. The user-selected filtered papers, together with the specified research focus and their intent, serve as the contextual input for the subsequent bi-directional analysis engine.

\subsubsection{Bi-directional Analysis}
As shown in \autoref{fig:backend}, \rh{the key to generate each narrative perspective is to identify its core components (see \autoref{fig:perspective}): a contribution statement, thematic clusters, and the individual papers within each cluster. Our solution leverages LLMs (OpenAI’s o3-mini) for their strong natural language understanding and reasoning capabilities to generate diverse narrative perspectives and their constituent components. }

\textbf{Narrative Schema: A Unified Schema for Data Flow and Mixed-Initiative Interaction.}
\rh{To support both top-down and bottom-up exploration, we designed a structured JSON schema (see \autoref{fig:backend}-iii) to represent the key components of a candidate narrative perspective. Specifically, the schema includes:}

\begin{itemize}
    \item  a particular clustering of papers, \texttt{paper\_assign}
    \item  the extracted themes of these clusters, \texttt{cluster\_theme}
    \item  a synthesized contribution statement that captures the unique framing of this perspective, \texttt{contribution\_statement}
\end{itemize}

\rh{As the central data layer of our system, this schema can be populated via LLM-generated outputs (top-down) or directly edited by users (bottom-up). It thus serves as a consistent and manipulable representation of narrative components, enabling seamless transitions between automated generation and human-driven refinement. This schema also underpins our prompt chain design, which we detail below in both top-down and bottom-up workflows.}

\textbf{Top-Down Reasoning: Framework-guided Clustering and Theme Synthesis.}
\rh{In the top-down workflow, the prompt chain includes two modules of context and instructions before the JSON output restrictions (see \autoref{appendix1} for the full prompt). The first module is a narrative structure-specific instruction module (see \autoref{fig:backend}-i). In which, we translated each narrative framework into natural language, defining a) the concepts of research statement, paper clusters, and themes; b) the hierarchy from statement, to clusters and individual papers; and c) semantic relations among clusters (\eg sequential in linear, orthogonal in coordinate). The second module is a reasoning chain guiding the LLMs through three steps (see \autoref{fig:backend}-ii): identifying distinctive features among papers, abstracting cluster themes, synthesizing potential contribution statements.} The output of this process is the structured JSON schema storing a set of candidate narrative perspectives and their components (see \autoref{fig:backend}-iii).

\textbf{Bottom-Up Reasoning: User-driven Re-Clustering and Narrative Refinement.}
In the bottom-up workflow, users interact with the Narrative Schema through the frontend interface to re-group papers, refine cluster themes, or revise contribution statements. Specifically, users can selectively lock components to preserve their edits, such as \texttt{paper\_assign} (by manually re-grouping papers), \texttt{cluster\_theme} (by locking thematic cluster boxes), or \texttt{contribution\_statement} (by locking the statement box). \rh{Once an update is requested, the LLM updates only the unlocked components (see \autoref{appendix2} for the full prompt).}

For example, in \autoref{fig:backend}-iv, a user groups the papers first, locks the \texttt{paper\_assign}, and requests updates from \toole. Upon request, \tool updates the remaining components (\ie contribution statement and cluster themes in this case) based on the locked parts. The analysis yields corresponding \texttt{cluster\_theme} and \texttt{contribution\_statement}.

\subsubsection{Evaluation and Ranking}
Before presenting the candidate perspectives to users, we evaluate each LLM-generated narrative perspective using SentenceTransformer-\texttt{all-MiniLM-L6-v2}\footnote{See \url{https://huggingface.co/sentence-transformers/all-MiniLM-L6-v2}}. The system computes a \textit{Final Score} by aggregating five equally weighted metrics that assess both semantic coherence and structural alignment. The first four metrics focus on semantic relationships: Statement-Cluster Alignment (\texttt{SCA}) evaluates semantic alignment between clusters and their parent statements; Intra-Cluster Cohesion (\texttt{ICC}) quantifies internal cluster consistency; Adjusted Rand Index (\texttt{ARI}) measures cluster coherence against unsupervised clustering results; and Paper-Cluster Similarity (\texttt{PCS}) verifies paper alignment with assigned cluster themes.

The fifth metric, Structural Consistency (\texttt{SC}), ensures that generated structures align with their intended narrative frameworks through pattern-specific evaluations. For Parallel patterns, we compute Cluster Separability to ensure distinct thematic boundaries between clusters. In Linear patterns, we assess Cluster Orderedness to verify logical progression between consecutive clusters. For Circular patterns, we measure Cluster Interdependence to examine mutual relationships among all clusters. In Coordinate patterns, we evaluate Axis Separability to confirm the proper distribution of clusters along different axes.

The final score aggregates these metrics as: \(\text{FinalScore} = 0.2 \cdot \texttt{SCA} + 0.2 \cdot \texttt{SC} + 0.2 \cdot \texttt{ARI} + 0.2 \cdot \texttt{PCS} + 0.2 \cdot \texttt{ICC} \). Each metric is normalized to [0, 1], and the four highest-scoring perspectives (the number is based on the findings of our pilot study, see \autoref{pilot}) are presented to users through the interface.

\subsubsection{Slide Generation}
The final backend module generates presentation slides from user-confirmed structures using the Python PPTX package\footnote{See \url{https://pypi.org/project/python-pptx/} for documentation}. We developed four templates aligned with four narrative frameworks \footnote{\rh{See supplementary for four slide templates}}, enabling export to PowerPoint for further refinement. In addition, we utilized Recraft\footnote{See \url{https://www.recraft.ai/docs} for documentation} to automatically generate images corresponding to each narrative perspective, aligning the visuals with the narration and enhancing overall engagement.

The system's frontend is implemented in Vue\footnote{See \url{https://vuejs.org/}}, with a Flask\footnote{See \url{https://flask.palletsprojects.com/en/stable/}} backend. Asynchronous data exchange facilitates seamless communication, while D3.js\footnote{See \url{https://d3js.org/} for documentation} powers interactive tree chart visualization corresponding to the four frameworks with real-time updates.
\section{Evaluation}

To evaluate the \toole, we conducted a user study with 12 HCI researchers. Specifically, our evaluation aims to investigate: 1) the usability and overall experience of \toole, \ie how useful and easy it is to use \toole, 2) if and how \tool supported researchers’ exploration of diverse research narratives.

\begin{figure*}[t]
  \centering
  \includegraphics[width=\linewidth]{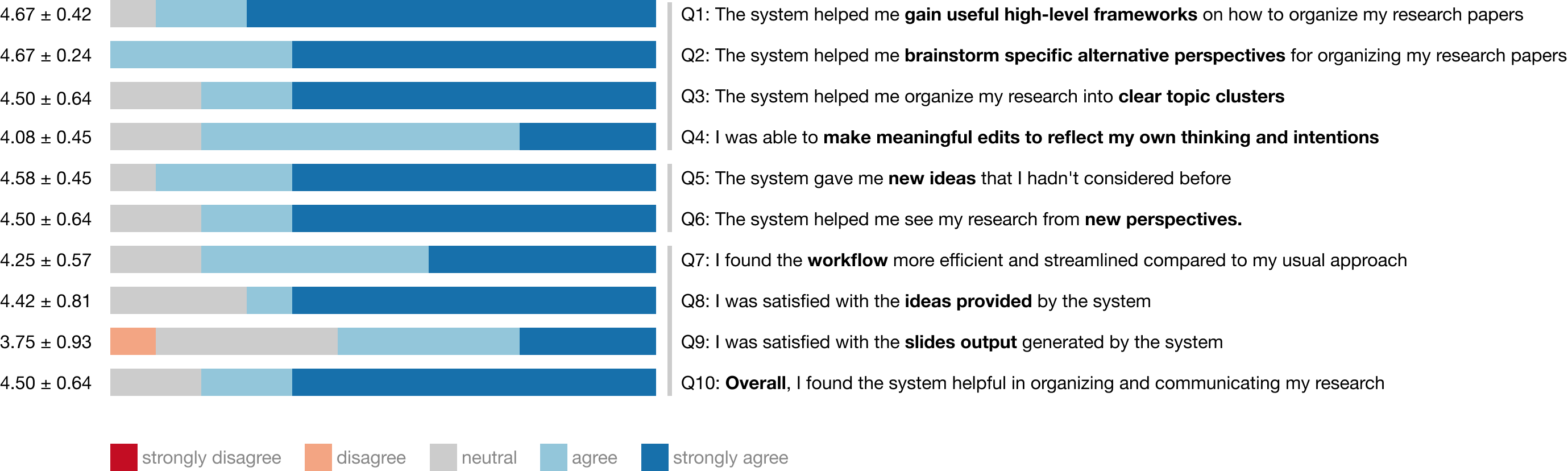}
  \caption{Assessment of participants' perception of \tool in terms of \textit{exploration} (Q1-Q4), \textit{inspiration} (Q5-Q6), and \textit{satisfaction} (Q7-Q10).}
  \label{fig:results}
\end{figure*}

\subsection{Methods} 
\label{pilot}
Before our main user study, we performed a pilot study with 4 participants to refine the study procedures. A key insight from this pilot study concerned the number of sparks (narrative perspectives) to provide under each framework. Participants indicated that 3-5 sparks struck an appropriate balance in terms of cognitive workload. Based on this feedback, we standardized the number of sparks to 4 per framework in our main user study to ensure a manageable yet sufficient workload for participants.

\subsubsection{Participants}
We recruited 12 researchers (P1-P12) through an open call on social media (4 female, 8 male, other gender options were provided). They span various career stages from senior PhD students to postdoctoral researchers and a research assistant professor. The participants had substantial publication records, ranging from 8 to 26 publications. Their research expertise covered a diverse range of HCI subfields, including virtual and augmented reality, visual analytics, accessibility, computer graphics, affective computing, cultural heritage, information visualization, human-AI collaboration, and online communities. This diversity in research backgrounds ensured that our evaluation captured perspectives from different areas within the HCI community.

\subsubsection{Procedures and Data Collection} 
Our user study comprised four stages: (1) a brief introduction; (2) a pre-interview before interaction with \toole; (3) free exploration with \toole; and (4) a post-interview including questionnaires. The whole procedure lasts around 1 to 1.5 hours. Details are as follows:

\textbf{Brief Introduction}. The study began with obtaining participant consent and introducing the project. Participants were presented with their primary task: organizing publications for a hypothetical job talk. Specifically, the task required them to re-organize their publications from perspectives different from their previous approach. Using \tool to assist in this process, they needed to draft at least two pages of slides. Time was allocated for participants to consider their potential audience and their intentions for highlighting their research. 

\textbf{Pre-interview and Brainstorming}. Participants then engaged in a pre-interview where they described their preliminary thoughts on organizing and presenting their body of work, along with their justifications. Pen and paper were provided for participants to illustrate their ideas. The facilitators asked follow-up questions when clarification was needed. This phase was crucial for establishing a baseline understanding of participants' organizational thinking before system interaction. The pre-interview was audio recorded. 

\textbf{Free Exploration with \toole}. After the pre-interview, we provided a 2-minute tutorial on \tool and allowed participants 5 minutes for initial exploration. Participants then interacted with \tool freely without time constraints. During the interaction, the laptop's screen was recorded. 

\textbf{Post-interview and Questionnaires}. Upon completing their interaction with \toole, participants completed questionnaires about their experience and then engaged in a post-interview. Specifically, the data collected included: 
\begin{itemize}
    \item \textbf{Quantitative:}
    \begin{itemize}
        \item \textit{Overall usability:} System Usability Scale (SUS) \cite{brooke2013sus} using a 5-point Likert scale.
        \item \textit{Workload:} NASA Task Load Index (NASA-TLX) \cite{hart1988development}, using a 7-point Likert scale.
        \item \textit{User experience questionnaire:} A 10-question survey using 5-point Likert scales, evaluating participants' perceptions of (1) the exploration process (inspired by \cite{wonderflow, vistalk}), (2) inspiration (inspired by \cite{suh2024luminate, chou2023talestream}), and (3) overall satisfaction. The specific questions can be found in \autoref{fig:results}.
        \item \textit{Ratings on generated narrative perspectives:} Participants were asked to rate each spark (16 in total) on a 5-point scale based on its reasonableness and helpfulness. Additionally, they were asked to indicate which sparks they had previously considered, allowing us to distinguish between novel ideas generated by \tool and those already conceived by participants.
    \end{itemize}

    \item \textbf{Qualitative:}
    \begin{itemize}
        \item We conducted semi-structured interviews to collect participants' thoughts and feedback during the exploration process. Sample questions included: ``\textit{Please describe your overall experiences with \toole},'' ``\textit{What are your opinions on the frameworks and sparks provided by the system?},'' ``\textit{Did you find new perspectives to frame your research, and how did our system support this?},'' and ``\textit{What suggestions do you have for improving the system?}''
    \end{itemize}
\end{itemize}

\subsection{Findings 1: Overall Usability and User Experiences}
\subsubsection{Usability and Workload}
The SUS and NASA-TLX measurements indicate that \tool is easy to use and imposes a low cognitive burden, while supporting high levels of task success during exploratory activities. Specifically, the System Usability Scale yielded an average score of 84.38 (SD = 19.54), which falls within the range of excellent usability \cite{brooke2013sus}. In terms of workload (measured on a 7-point scale), participants reported low mental demand (M = 3.08, SD = 2.07), very low physical demand (M = 1.42, SD = 0.67), and low time pressure (M = 2.33, SD = 1.67) while using \toole. Notably, participants reported a high sense of success in completing the task (M = 5.50, SD = 1.31), suggesting that the system effectively supported their goals. Reported effort levels were moderate (M = 3.42, SD = 1.51), and frustration remained low (M = 1.75, SD = 0.97).

\subsubsection{Overall User Experience}
As shown in \autoref{fig:results}, results from the user experience questionnaire (N = 12, 5-point Likert scale) indicate that participants were generally satisfied with \tool (M = 4.39, SD = 0.59). Specifically, participants gave positive evaluations across three key dimensions. The \textit{exploration} process dimension (4 questions) received high ratings (M = 4.48, SD = 0.52), indicating that participants found the system effective for organizing research and brainstorming alternative perspectives. The \textit{inspiration} dimension (2 questions) received the highest ratings (M = 4.54, SD = 0.69), suggesting that \tool helped users discover new ideas and view their work from different angles. The \textit{satisfaction} dimension (4 questions) was also rated positively (M = 4.23, SD = 0.73), with slide output satisfaction showing the most room for improvement.

\subsection{Findings 2: Co-Exploration Processes with \toole} 

The analysis of post-interviews and participants' ratings on the generated narrative perspectives, complemented by in-situ observations during the study, revealed how \toole: (1) enhanced the efficiency of narrative ideation; (2) supported the co-exploration through both top-down and bottom-up approaches; and (3) inspired reflections on participants’ research corpus beyond the immediate task of narrative construction. These findings demonstrated that \tool was helpful in exploring alternative research narratives, and also surfaced insights about participants’ further expectations.

\subsubsection{Facilitating Efficient Narrative Ideation}
All participants reported that \tool improved the efficiency of narrative exploration by offering diverse options or perspectives for them. This efficiency emerged from multiple system affordances:

First, most participants noted that both the frameworks and narrative perspectives provided by \tool \textbf{offered a wide range of starting points} for exploration. Frameworks gave access to high-level, abstract structures, while perspectives under each framework served as more concrete, interpretable sparks. As P7 described: ``\textit{It's really helpful to have these frameworks available. It makes it much easier to try different structures and think about how to organize my publications.}'' When it came to narrative perspectives, participants highlighted how they helped reduce the barrier to starting narrative ideation. P8 remarked: ``\textit{I probably couldn't have come up with so many different approaches on my own. Having these perspectives presented to me made the task of creating my research narrative feel less daunting.}'' 

In addition to the rich and diverse suggestions provided, participants appreciated the \toole's \textbf{clear workflow that streamlined their exploration}. We found that the top-down exploration helped participants structure their thoughts more systematically. P9 shared: ``\textit{I like its ability to guide me through a step-by-step exploration of different narrative aspects; when I think on my own, my thoughts tend to be scattered.}'' At the same time, participants mentioned \tool facilitated their bottom-up exploration with real-time updates. P9 emphasized: ``\textit{Grouping my papers and testing different ideas with \tool is more efficient than sketching on paper. I can quickly get feedback on how these papers might work together when organized in different ways.}''

Finally, the \toole’s \textbf{ability to output slide drafts} was also appreciated by participants, as these drafts helped them transition more directly into discussion and collaboration. As P6 noted: ``\textit{With these slides, I can more quickly take the materials for discussion with others; previously, I had to make separate slides from the beginning after organizing my thoughts.}'' Beyond immediate use, some participants imagined integrating these outputs into longer-term academic assets. For instance, P12 suggested: ``\textit{With some optimization, I think I will put the visuals on my personal website.}''

\subsubsection{Co-Exploration Process} 
Beyond improving efficiency, our findings revealed that \tool supported an exploratory and reflective process for crafting research narratives. This process began with trust building, human-AI co-exploration of narrative perspectives, and participants’ reflection on their research beyond the immediate task of narrative construction.

\textbf{Initial Trust Establishment.} We noticed the participants commonly began by quickly scanning among the frameworks and sparks. A sense of initial trust was often formed when they recognized familiar or sensible ideas within the system’s suggestions. As P2 explained: ``\textit{When I saw ideas I had thought of before, I felt the system worked well; it made me feel that the sparks it provided were logical and reasonable.}''

\textbf{Co-Exploration of Narrative Frameworks.} 
Participants often initiated deeper exploration by examining high-level narrative frameworks. The presence of multiple frameworks sparked curiosity and motivated participants to explore the perspectives under each of them. As P7 reflected: ``\textit{These frameworks seemed reasonable, making me want to explore each of them.}'' 

\textbf{Co-Exploration of Diverse Narrative Perspectives.} 
When exploring the sparks (i.e., narrative perspectives) within a specific framework, the \textbf{discovery of novel perspectives} often facilitated deeper engagement with the exploration process. For example, P1 noted: ``\textit{I designed and observed different user behavior in VR. However, I never thought to categorize my work through the lens of ‘perception.’ It surprised me and made me want to explore more.}'' 

Besides, we found that \textbf{keywords} associated with contribution statements and thematic clusters played a critical role. The keywords generated by \tool offered diverse entry points and remained open to interpretation, allowing participants to project their own understanding. As P6 noted: ``\textit{Seeing these keywords, I will interpret them first before I check the details in the workspace.}'' We also observed that participants tended to prefer abstract or conceptual keywords over specific ones, as the former provided more room for interpretation. As P4 commented: ``\textit{There were two cluster names—one was ‘tutorial-guided interaction’ and the other was ‘hand gesture interaction.’ The former left room for imagination, while the latter felt too specific and less suitable.}''

Additionally, the interpretive space created by these keywords often inspired participants to \textbf{shift into bottom-up exploration} to test and refine their own ideas. For example, P12 mentioned: ``\textit{When I saw the keywords ‘decision-making process,’ I got an aha moment. But I didn't like the way it grouped my papers. So, I tried to fix the contribution statement, assign the papers myself, and request the system to return new cluster themes.}''

\textbf{Reflections Emerged During the Exploration.}
We noticed that the exploration process was a process to help researchers re-think their research through different perspectives. We also noticed that, in addition to the diverse candidates that served as the reflection trigger, \toole’s visual and conceptual organization prompted participants to think more broadly about their research positioning and future directions. Specifically, the visual layout of the workspace enabled participants to identify latent patterns and gaps across their body of work. P4 shared: ``\textit{When it showed three clusters describing how my work innovates tangible interface input methods—gesture input, object input, and spatial context input—I noticed I had fewer papers in the last cluster, suggesting a potential direction for future research.}''  

\subsection{Findings 3: Challenges and Tensions Observed}
Since participants are most familiar with the content and context of their own publications, we consider their feedback particularly valuable. We summarize their expectations as part of our findings to surface the tensions and challenges involved in supporting the reframing of one’s own intellectual work.

\subsubsection{Complexities in How Users Perceive and Rate Sparks}
We collected a total of 192 spark ratings (4 sparks × 4 narrative structures × 12 participants), yielding an overall mean score of 3.18 (SD = 1.29). To further examine how familiarity shaped evaluations, we compared ratings of sparks that captured ideas participants had already considered (“thought‑of,” N = 66) versus those that were entirely new (“not‑thought‑of,” N = 126). A paired-samples \textit{t}-test showed that “thought‑of” sparks were rated slightly higher (M = 3.41, SD = 1.29) than “not‑thought‑of” sparks (M = 3.06, SD = 1.27), though the effect was only marginal, \textit{t}(11) = 2.14, \textit{p} = .056.

We noticed an inconsistency between the average spark scores and the otherwise positive quantitative and qualitative data. Despite a moderate mean rating, the SUS scores, user experience measures, and participants’ interview responses indicated that \tool was effective in helping users explore diverse research narratives. This discrepancy calls for a more nuanced interpretation of the spark ratings. 

Specifically, we observed that spark scores were sensitive to the characteristics of users’ inputs, particularly the number of papers and the degree of thematic coherence among them. For example, P11 selected four papers, two of which were topically disjointed. P11 gave lower ratings to most sparks, with only one spark receiving a score of 4. Reflecting on this, the participant acknowledged, ``\textit{I know my publications were not that coherent, which made me headache. But at least, your system yielded one helpful perspective for me called `technological mediation in the interaction'.}'' This was the one spark that received a rating of 4. Drawing from this case, we interpret that the spark scores do not fully capture the extent to which participants were inspired by \toole, as a smaller number of papers may limit the system’s ability to form meaningful clusters from the input, while greater topical diversity can reduce the semantic coherence required to generate compelling sparks. In other words, both factors may constrain the quality of the generated sparks.

\subsubsection{Expectations for More Flexible Framework Organization}
The exploration of different narrative frameworks made participants express interest in more open-ended and customizable structural frameworks. They wished for the ability to go beyond predefined templates and experiment with representations that better reflect the nuances of their work. For instance, P3 remarked: ``\textit{I wish I could arrange and combine these frameworks arbitrarily. For instance, I want to add two additional nodes to one edge of a circular structure and see what happens.}'' P6 emphasized the value of aligning frameworks with the structure of specific theoretical models: ``\textit{Different theories have different levels, and these levels and nodes could potentially resonate with one’s papers.}'' Meanwhile, both P1 and P5 pointed out the importance of relative positioning in coordinate-based structures. As P5 explained: ``\textit{I know that this paper (pointing at the screen), the degree of AI automation is higher, so it should be placed closer to the extreme of this axis.}'' These insights indicate a strong desire for more granular, theory-aligned, and manipulable narrative structures. 
\section{Discussion}
In this paper, we targeted HCI researchers' own challenges in exploring alternative research narratives, and presented \tool to support this process. In this section, we reflect on the key design decisions and associated trade-offs, and discuss implications and directions for future research.

\subsection{Supporting Academic Storytelling and Communication}

By targeting the challenge of helping HCI researchers construct alternative narratives around their own body of work, our study contributes to a growing line of research on how interactive systems can support academic practice \cite{van2014citnetexplorer,hyland2018narrative,Heer}. While many existing tools focus on literature discovery \cite{sultanum2020understanding,kang2023synergi,Xiao2023a}, writing mechanics \cite{nguyen2024human,Shao2024a}, or citation management \cite{sultanum2020understanding}, our work centers on a distinct yet critical scenario: \textit{\textbf{academic storytelling and communication}}. \rh{We approach this direction by examining and supporting research narratives constructions based on one's own publications. In the following, we reflect on our work and offer insights that may inform future research in this emerging space.}

\subsubsection{\rh{Narrative Frameworks as Design Patterns and Conceptual Scaffolds for Academic Storytelling}}
\rh{Our investigation started from identifying narrative frameworks (\ie \textit{parallel}, \textit{linear}, \textit{coordinate}, and \textit{circular}). While the four structures are not exhaustive, we argue that they may serve as design patterns~\cite{Bach2023, 10.1145/3613904.3642754} as well as conceptual scaffolds~\cite{Setlur2020, 10.1145/3406522.3446012} in guiding future research and system development.} 

\rh{\textbf{\textit{As design patterns}}, these frameworks offer reusable structural templates for organizing publications into narrative perspectives. They provide a foundation for interface designs that support narrative construction, and can be extended to accommodate greater flexibility. For example, future systems might allow users to adjust the number of clusters, combine multiple structural logics, or interactively manipulate narrative layouts to better reflect individual goals and preferences. However, it is worth noting that increased flexibility may also introduce new challenges. For instance, over-fragmentation could compromise narrative coherence, highlighting the need for additional design considerations to support manageable and meaningful structure selection.}  \rh{\textbf{\textit{As conceptual scaffolds}}, these frameworks offer cognitive support for complex narrative construction. They help users interpret and decompose ideas, enabling them to make sense of the bodies of their own work. Beyond academic storytelling, such scaffolds may generalize to other interpretive tasks that require organizing multiple components into a coherent whole, such as structuring course modules planning~\cite{10.1145/3654777.3676390}.} 

\subsubsection{\rh{Academic Storytelling as an Exploratory and Reflective Process}}
\rh{The evaluation of \tool showed that researchers appreciated the support in academic storytelling, and revealed that engaging with \tool not only improved efficiency but also fostered a co-exploration process that encouraged reinterpretation and reflection on their research. These observations suggest that academic storytelling can be understood as \textbf{\textit{an exploratory and reflective activity}}, rather than merely a task of communication. In light of this,} we call for greater attention to the design of systems that support researchers in exploring and reflecting on their own intellectual trajectories. \rh{In our study, we observed that using keywords as outputs created space for reflection. This \textbf{\textit{lightweight and suggestive design}} encourages users to reconsider and reframe their own work, rather than passively accept system-generated content. Such an approach offers a promising direction for future systems that aim to balance guidance and interpretive flexibility.} Although our study focused on HCI researchers, the need to make sense of one’s academic trajectory exists across disciplines and contexts, such as grant writing, portfolio curation, or job applications. We see this as an emerging space for academic and professional storytelling that is worth further exploration.

\subsubsection{\rh{Challenges in Supporting Narrative Construction with Familiar Content}}
Throughout our study, we identified several design challenges specific to the academic storytelling and communication scenario. Unlike tasks involving unfamiliar content~\cite{datadirector,dataplayer}, participants were deeply familiar with the material, as they were working with their own publications. This familiarity introduced a unique tension: participants wanted the system to inspire them with new ideas, but also expected its suggestions to align with their self-understanding. As a result, they tended to be more critical of the system’s outputs, \rh{and their assessments were heavily influenced by subjective expectations}. For example, spark ratings were slightly higher for ideas they had already considered, suggesting a preference for outputs that resonated with their existing thinking. To support such scenarios, future systems could consider providing users with more agency in the interaction~\cite{Heer2019, Instructions}. \rh{This may include exposing parts of the system’s reasoning process or enabling interaction with scaffolded reasoning steps~\cite{10.1145/3706598.3714135}, in order to provide transparency that fosters trust and supports more reflective evaluation during research narrative construction.}

\subsection{Designing Human-AI Collaboration for Mixed-Initiative Interaction}
To support the exploratory nature of research narrative construction, \tool enables human-AI co-exploration through both top-down and bottom-up approaches. \rh{To this end, we followed established practices in guiding LLMs by translating narrative construction practices into structured prompt instructions~\cite{10.1145/3706598.3713726}. To further support mixed-initiative interaction around the generated narrative components, we \textbf{\textit{designed the LLM outputs as shared, editable representations}} that bridge model reasoning and user refinement. In the following, we reflect on the design of these editable representations and discuss their broader generalizability.}

\subsubsection{\rh{Editable Representations for Mixed-Initiative Interaction}}
\rh{In our current work, the shared, editable representations are implemented using a structured JSON schema, which defines the key elements of a narrative perspective, including the contribution statement, thematic clusters, and paper assignments. By making these fields editable and accessible across both the frontend and backend, the system allows users to iteratively adjust, reframe, and reorganize the generated content~\cite{dataplaywright, NotePlayer}. In addition to \textbf{\textit{enabling mixed-initiative interaction}}, this schema design also \textbf{\textit{enhances the extensibility}} of \toole. For example, new narrative frameworks can be easily incorporated by defining structure-specific prompts and mapping their outputs to the same schema fields, without altering the overall system architecture.} 

\rh{Beyond narrative construction, the concept of editable representations offers an alternative solution for other interpretive tasks that involve LLMs reasoning and human refinement. This design \textbf{\textit{reframes a generative task as an interactive process of interpretation and reconstruction}} between human and AI. Based on our findings, decomposing narrative construction into modular components helps users avoid the cognitive burden of starting from scratch and provides concrete entry points for reflection. This design principle may be generalized to other domains, including literature surveys~\cite{Xiao2023a}, idea brainstorming~\cite{Liu2024b}, Socratic-style learning~\cite{Liu2024e}, and creativity support tools~\cite{Angert2023, DVSurvey}. While our current interface supports text-based refinement, the underlying concept aligns with recent visions of dynamic abstractions~\cite{10.1145/3672539.3686706, 10.1145/3706598.3713723}, which advocate for interpretable and manipulable intermediate representations. Future systems may build on this approach by designing multi-modal interactions, such as sketch- or diagram-based interfaces.}

\subsubsection{\rh{Trade-offs in Input Scope for Narrative Generation}}
However, our use of LLMs also revealed challenges, particularly around how the scope of input influences the perceived quality of output. \rh{Our current approach, which relies on paper titles and abstracts, introduces a trade-off. On one hand, it reduces the input burden (\eg users do not need to upload full papers), and abstracts often capture the core ideas of a publication, supporting our goal of surfacing narrative inspirations. On the other hand, this limited input may constrain the specificity and depth of the generated clusters and contribution statements.} We observed that when users selected only a small number of papers, the system had fewer materials from which to extract meaningful patterns or construct abstract framings. At the same time, users often had highly specific interpretations of these papers, making them more sensitive to mismatches or inaccuracies in the system's suggestions. Although this tension was amplified by the limited input scope, we believe the underlying challenge would persist even with full-text access. Researchers with more publications naturally provide a richer foundation for the system to identify various themes. Addressing this challenge may require models or systems with more adaptive reasoning capabilities. Such systems would need to operate effectively across varying input sizes and tailor the level of abstraction and suggestion to the quantity and specificity of the available content.

\subsection{Limitations and Future Work}
We acknowledge several limitations of our current work and would like to outline potential directions for future research. \rh{First, our current study focused on HCI researchers, and both the narrative frameworks and user study findings primarily reflect experiences from within the HCI community. The participants in our formative study were mainly early-career researchers preparing for the job market. Although the insights from the formative study and content analysis were discussed with senior HCI researchers, the findings have yet to be examined more broadly across the HCI community. A natural next step is to extend \tool to a more diverse group of HCI researchers, as well as to other computer science subfields, in order to explore both shared narrative patterns and domain-specific differences in research narrative construction.}

\rh{Second, the four narrative frameworks are not intended to be exhaustive. Future work may explore additional frameworks or investigate more flexible, customizable structures to support personalized and diverse forms of narrative construction. Third, although \tool goes beyond standard chat-based interfaces by embedding domain knowledge and providing a structured workflow, our current study did not include a baseline comparison. A valuable direction for future research is to evaluate how \tool performs in comparison with general-purpose tools like ChatGPT, particularly when both are equipped with narrative knowledge.

Finally, in terms of input for LLMs, we currently rely on titles and abstracts retrieved from Google Scholar to reduce user effort}. Future work could incorporate multi-modal inputs uploaded by users, such as PDFs, slides, or project documents, to support deeper narrative analysis and more diverse output formats. Additionally, beyond the published papers, the ongoing projects are valuable resources to be considered. 
\section{Conclusion}

In this paper, we targeted the challenges HCI researchers face in constructing alternative research narratives. Building on empirical findings from a formative study and content analysis, we developed \toole, a human-AI co-exploration system that supports narrative construction through both top-down and bottom-up approaches. Powered by a bi-directional analysis engine, \tool enables users to generate, revise, and reinterpret their research narratives from diverse perspectives. Our user study with 12 HCI researchers demonstrated that \tool not only improved the efficiency of narrative construction but also supported an exploratory and reflective process. Based on these findings, we offer implications for future research on 1) academic storytelling and communication and 2) designing human-AI collaborative systems for mixed-initiative interaction.

\begin{acks}
This work was partially supported by RGC GRF grant 16210722, and was supported in part by the EdUHK-HKUST Joint Centre for Artificial Intelligence (JC\_AI) research scheme: Grant No. FB454. We sincerely thank all participants for their time and efforts, and the anonymous reviewers for their constructive feedback, which significantly improved the quality of this paper.

We would like to thank Dr. Chengbo Zheng and Liwenhan Xie for their valuable discussions during the revision of this paper. We also thank Junze Li, Dr. Yao Wang, Hanlu Ma, Dr. Linping Yuan, Dingdong Liu, and Zheng Wei for their insightful comments on earlier drafts, and Jiawen Liu and Meiying Li for their help with proofreading. Finally, the first author wishes to express special thanks to Dr. Yulin Tian, Ziqi Pan, Qiaoyi Chen, and to those who showed patience and support during a personally challenging time. Their encouragement and willingness to simply be there made it possible to carry this work through to completion.

\end{acks}

\balance

\bibliographystyle{ACM-Reference-Format}
\bibliography{main} 


\begin{thebibliography}{74}

%%% the Web address.

\ifx \showCODEN    \undefined \def \showCODEN     #1{\unskip}     \fi
\ifx \showDOI      \undefined \def \showDOI       #1{#1}\fi
\ifx \showISBNx    \undefined \def \showISBNx     #1{\unskip}     \fi
\ifx \showISBNxiii \undefined \def \showISBNxiii  #1{\unskip}     \fi
\ifx \showISSN     \undefined \def \showISSN      #1{\unskip}     \fi
\ifx \showLCCN     \undefined \def \showLCCN      #1{\unskip}     \fi
\ifx \shownote     \undefined \def \shownote      #1{#1}          \fi
\ifx \showarticletitle \undefined \def \showarticletitle #1{#1}   \fi
\ifx \showURL      \undefined \def \showURL       {\relax}        \fi
% invisible to TeX
\providecommand\bibfield[2]{#2}
\providecommand\bibinfo[2]{#2}
\providecommand\natexlab[1]{#1}
\providecommand\showeprint[2][]{arXiv:#2}

\bibitem[Aitken(2010)]%
        {aitken2010becoming}
\bibfield{author}{\bibinfo{person}{Avril Aitken}.} \bibinfo{year}{2010}\natexlab{}.
\newblock \showarticletitle{Becoming an academic: Professional identity on the road to tenure}.
\newblock \bibinfo{journal}{\emph{The Journal of Educational Thought (JET)/Revue de la Pens{\'e}e {\'E}ducative}} (\bibinfo{year}{2010}), \bibinfo{pages}{55--68}.
\newblock


\bibitem[Angert et~al\mbox{.}(2023)]%
        {Angert2023}
\bibfield{author}{\bibinfo{person}{Tyler Angert}, \bibinfo{person}{Miroslav Suzara}, \bibinfo{person}{Jenny Han}, \bibinfo{person}{Christopher Pondoc}, {and} \bibinfo{person}{Hariharan Subramonyam}.} \bibinfo{year}{2023}\natexlab{}.
\newblock \showarticletitle{{Spellburst: A Node-based Interface for Exploratory Creative Coding with Natural Language Prompts}}. In \bibinfo{booktitle}{\emph{Proceedings of the 36th Annual ACM Symposium on User Interface Software and Technology}}. \bibinfo{publisher}{ACM}, \bibinfo{pages}{1--22}.
\newblock


\bibitem[Aristotle et~al\mbox{.}(1954)]%
        {aristotle1954rhetoric}
\bibfield{author}{\bibinfo{person}{Aristotle Aristotle} {et~al\mbox{.}}} \bibinfo{year}{1954}\natexlab{}.
\newblock \bibinfo{booktitle}{\emph{Rhetoric}}. Vol.~\bibinfo{volume}{11}.
\newblock \bibinfo{publisher}{Modern Library New York}.
\newblock


\bibitem[Bach et~al\mbox{.}(2023)]%
        {Bach2023}
\bibfield{author}{\bibinfo{person}{Benjamin Bach}, \bibinfo{person}{Euan Freeman}, \bibinfo{person}{Alfie Abdul-Rahman}, \bibinfo{person}{Cagatay Turkay}, \bibinfo{person}{Saiful Khan}, \bibinfo{person}{Yulei Fan}, {and} \bibinfo{person}{Min Chen}.} \bibinfo{year}{2023}\natexlab{}.
\newblock \showarticletitle{{Dashboard Design Patterns}}.
\newblock \bibinfo{journal}{\emph{IEEE Transactions on Visualization and Computer Graphics}} \bibinfo{volume}{29}, \bibinfo{number}{1} (\bibinfo{year}{2023}), \bibinfo{pages}{342--352}.
\newblock
\showISSN{19410506}


\bibitem[Bravo et~al\mbox{.}(2020)]%
        {bravo2020researcher}
\bibfield{author}{\bibinfo{person}{Maricela Bravo}, \bibinfo{person}{Jos{\'e}~A Reyes-Ortiz}, {and} \bibinfo{person}{Isabel Cruz}.} \bibinfo{year}{2020}\natexlab{}.
\newblock \showarticletitle{Researcher profile ontology for academic environment}. In \bibinfo{booktitle}{\emph{Advances in Computer Vision: Proceedings of the 2019 Computer Vision Conference (CVC), Volume 1 1}}. Springer, \bibinfo{pages}{799--817}.
\newblock


\bibitem[Brooke(2013)]%
        {brooke2013sus}
\bibfield{author}{\bibinfo{person}{John Brooke}.} \bibinfo{year}{2013}\natexlab{}.
\newblock \showarticletitle{SUS: a retrospective.}
\newblock \bibinfo{journal}{\emph{Journal of usability studies}} \bibinfo{volume}{8}, \bibinfo{number}{2} (\bibinfo{year}{2013}).
\newblock


\bibitem[C\^{a}mara et~al\mbox{.}(2021)]%
        {10.1145/3406522.3446012}
\bibfield{author}{\bibinfo{person}{Arthur C\^{a}mara}, \bibinfo{person}{Nirmal Roy}, \bibinfo{person}{David Maxwell}, {and} \bibinfo{person}{Claudia Hauff}.} \bibinfo{year}{2021}\natexlab{}.
\newblock \showarticletitle{Searching to Learn with Instructional Scaffolding}. In \bibinfo{booktitle}{\emph{Proceedings of the 2021 Conference on Human Information Interaction and Retrieval}} (Canberra ACT, Australia) \emph{(\bibinfo{series}{CHIIR '21})}. \bibinfo{publisher}{Association for Computing Machinery}, \bibinfo{address}{New York, NY, USA}, \bibinfo{pages}{209–218}.
\newblock
\showISBNx{9781450380553}


\bibitem[Capel and Brereton(2023)]%
        {Capel2023a}
\bibfield{author}{\bibinfo{person}{Tara Capel} {and} \bibinfo{person}{Margot Brereton}.} \bibinfo{year}{2023}\natexlab{}.
\newblock \showarticletitle{{What is Human-Centered about Human-Centered AI? A Map of the Research Landscape}}. In \bibinfo{booktitle}{\emph{Proceedings of the 2023 CHI Conference on Human Factors in Computing Systems, CHI'23}}. \bibinfo{publisher}{ACM}, \bibinfo{pages}{1--23}.
\newblock


\bibitem[Chang et~al\mbox{.}(2023)]%
        {chang2023citesee}
\bibfield{author}{\bibinfo{person}{Joseph~Chee Chang}, \bibinfo{person}{Amy~X Zhang}, \bibinfo{person}{Jonathan Bragg}, \bibinfo{person}{Andrew Head}, \bibinfo{person}{Kyle Lo}, \bibinfo{person}{Doug Downey}, {and} \bibinfo{person}{Daniel~S Weld}.} \bibinfo{year}{2023}\natexlab{}.
\newblock \showarticletitle{Citesee: Augmenting citations in scientific papers with persistent and personalized historical context}. In \bibinfo{booktitle}{\emph{Proceedings of the 2023 CHI Conference on Human Factors in Computing Systems}}. \bibinfo{pages}{1--15}.
\newblock


\bibitem[Chen et~al\mbox{.}(2025)]%
        {10.1145/3706598.3714135}
\bibfield{author}{\bibinfo{person}{Wei-Hao Chen}, \bibinfo{person}{Weixi Tong}, \bibinfo{person}{Ph.D. Case, Amanda}, {and} \bibinfo{person}{Tianyi Zhang}.} \bibinfo{year}{2025}\natexlab{}.
\newblock \showarticletitle{Dango: A Mixed-Initiative Data Wrangling System using Large Language Model}. In \bibinfo{booktitle}{\emph{Proceedings of the 2025 CHI Conference on Human Factors in Computing Systems}} \emph{(\bibinfo{series}{CHI '25})}. \bibinfo{publisher}{Association for Computing Machinery}, \bibinfo{address}{New York, NY, USA}, Article \bibinfo{articleno}{389}, \bibinfo{numpages}{28}~pages.
\newblock
\showISBNx{9798400713941}


\bibitem[Choe et~al\mbox{.}(2024)]%
        {choe2024fields}
\bibfield{author}{\bibinfo{person}{Kiroong Choe}, \bibinfo{person}{Eunhye Kim}, \bibinfo{person}{Sangwon Park}, {and} \bibinfo{person}{Jinwook Seo}.} \bibinfo{year}{2024}\natexlab{}.
\newblock \showarticletitle{Fields, Bridges, and Foundations: How Researchers Browse Citation Network Visualizations}. In \bibinfo{booktitle}{\emph{2024 IEEE Visualization and Visual Analytics (VIS)}}. IEEE, \bibinfo{pages}{146--150}.
\newblock


\bibitem[Chou et~al\mbox{.}(2023)]%
        {chou2023talestream}
\bibfield{author}{\bibinfo{person}{Jean-Pe{\"\i}c Chou}, \bibinfo{person}{Alexa~Fay Siu}, \bibinfo{person}{Nedim Lipka}, \bibinfo{person}{Ryan Rossi}, \bibinfo{person}{Franck Dernoncourt}, {and} \bibinfo{person}{Maneesh Agrawala}.} \bibinfo{year}{2023}\natexlab{}.
\newblock \showarticletitle{Talestream: Supporting story ideation with trope knowledge}. In \bibinfo{booktitle}{\emph{Proceedings of the 36th Annual ACM Symposium on User Interface Software and Technology}}. \bibinfo{pages}{1--12}.
\newblock


\bibitem[Cronon(1992)]%
        {cronon1992place}
\bibfield{author}{\bibinfo{person}{William Cronon}.} \bibinfo{year}{1992}\natexlab{}.
\newblock \showarticletitle{A place for stories: Nature, history, and narrative}.
\newblock \bibinfo{journal}{\emph{Journal of American history}} \bibinfo{volume}{78}, \bibinfo{number}{4} (\bibinfo{year}{1992}), \bibinfo{pages}{1347--1376}.
\newblock


\bibitem[De~Los~Reyes(2024)]%
        {de2024delivering}
\bibfield{author}{\bibinfo{person}{Andres De~Los~Reyes}.} \bibinfo{year}{2024}\natexlab{}.
\newblock \showarticletitle{Delivering Academic Job Talks as an Early Career Researcher}.
\newblock In \bibinfo{booktitle}{\emph{The Early Career Researcher's Toolbox: Insights into Mentors, Peer Review, and Landing a Faculty Job}}. \bibinfo{publisher}{Springer}, \bibinfo{pages}{117--122}.
\newblock


\bibitem[Dörk et~al\mbox{.}(2012)]%
        {6327277}
\bibfield{author}{\bibinfo{person}{Marian Dörk}, \bibinfo{person}{Nathalie Henry~Riche}, \bibinfo{person}{Gonzalo Ramos}, {and} \bibinfo{person}{Susan Dumais}.} \bibinfo{year}{2012}\natexlab{}.
\newblock \showarticletitle{PivotPaths: Strolling through Faceted Information Spaces}.
\newblock \bibinfo{journal}{\emph{IEEE Transactions on Visualization and Computer Graphics}} \bibinfo{volume}{18}, \bibinfo{number}{12} (\bibinfo{year}{2012}), \bibinfo{pages}{2709--2718}.
\newblock


\bibitem[Fan et~al\mbox{.}(2024)]%
        {10.1145/3654777.3676390}
\bibfield{author}{\bibinfo{person}{Haoxiang Fan}, \bibinfo{person}{Guanzheng Chen}, \bibinfo{person}{Xingbo Wang}, {and} \bibinfo{person}{Zhenhui Peng}.} \bibinfo{year}{2024}\natexlab{}.
\newblock \showarticletitle{LessonPlanner: Assisting Novice Teachers to Prepare Pedagogy-Driven Lesson Plans with Large Language Models}. In \bibinfo{booktitle}{\emph{Proceedings of the 37th Annual ACM Symposium on User Interface Software and Technology}} (Pittsburgh, PA, USA) \emph{(\bibinfo{series}{UIST '24})}. \bibinfo{publisher}{Association for Computing Machinery}, \bibinfo{address}{New York, NY, USA}, Article \bibinfo{articleno}{146}, \bibinfo{numpages}{20}~pages.
\newblock
\showISBNx{9798400706288}


\bibitem[Flowerdew and Wang(2015)]%
        {flowerdew2015identity}
\bibfield{author}{\bibinfo{person}{John Flowerdew} {and} \bibinfo{person}{Simon~Ho Wang}.} \bibinfo{year}{2015}\natexlab{}.
\newblock \showarticletitle{Identity in academic discourse}.
\newblock \bibinfo{journal}{\emph{Annual Review of Applied Linguistics}}  \bibinfo{volume}{35} (\bibinfo{year}{2015}), \bibinfo{pages}{81--99}.
\newblock


\bibitem[Hart and Staveland(1988)]%
        {hart1988development}
\bibfield{author}{\bibinfo{person}{Sandra~G Hart} {and} \bibinfo{person}{Lowell~E Staveland}.} \bibinfo{year}{1988}\natexlab{}.
\newblock \showarticletitle{Development of NASA-TLX (Task Load Index): Results of empirical and theoretical research}.
\newblock In \bibinfo{booktitle}{\emph{Advances in psychology}}. Vol.~\bibinfo{volume}{52}. \bibinfo{publisher}{Elsevier}, \bibinfo{pages}{139--183}.
\newblock


\bibitem[Hayatpur et~al\mbox{.}(2025)]%
        {10.1145/3706598.3713723}
\bibfield{author}{\bibinfo{person}{Devamardeep Hayatpur}, \bibinfo{person}{Brian Hempel}, \bibinfo{person}{Richard Lin}, {and} \bibinfo{person}{Haijun Xia}.} \bibinfo{year}{2025}\natexlab{}.
\newblock \showarticletitle{The Shapes of Abstraction in Data Structure Diagrams}. In \bibinfo{booktitle}{\emph{Proceedings of the 2025 CHI Conference on Human Factors in Computing Systems}} \emph{(\bibinfo{series}{CHI '25})}. \bibinfo{publisher}{Association for Computing Machinery}, \bibinfo{address}{New York, NY, USA}, Article \bibinfo{articleno}{883}, \bibinfo{numpages}{12}~pages.
\newblock
\showISBNx{9798400713941}


\bibitem[He et~al\mbox{.}(2019)]%
        {he2019paperpoles}
\bibfield{author}{\bibinfo{person}{Jiangen He}, \bibinfo{person}{Qing Ping}, \bibinfo{person}{Wen Lou}, {and} \bibinfo{person}{Chaomei Chen}.} \bibinfo{year}{2019}\natexlab{}.
\newblock \showarticletitle{PaperPoles: Facilitating adaptive visual exploration of scientific publications by citation links}.
\newblock \bibinfo{journal}{\emph{Journal of the Association for Information Science and Technology}} \bibinfo{volume}{70}, \bibinfo{number}{8} (\bibinfo{year}{2019}), \bibinfo{pages}{843--857}.
\newblock


\bibitem[Heer(2019)]%
        {Heer2019}
\bibfield{author}{\bibinfo{person}{Jeffrey Heer}.} \bibinfo{year}{2019}\natexlab{}.
\newblock \showarticletitle{{Agency plus automation: Designing artificial intelligence into interactive systems}}.
\newblock \bibinfo{journal}{\emph{Proceedings of the National Academy of Sciences of the United States of America}} \bibinfo{volume}{116}, \bibinfo{number}{6} (\bibinfo{year}{2019}), \bibinfo{pages}{1844--1850}.
\newblock


\bibitem[Heer et~al\mbox{.}(2023)]%
        {Heer}
\bibfield{author}{\bibinfo{person}{Jeffrey Heer}, \bibinfo{person}{Matthew Conlen}, \bibinfo{person}{Vishal Devireddy}, \bibinfo{person}{Tu Nguyen}, {and} \bibinfo{person}{Joshua Horowitz}.} \bibinfo{year}{2023}\natexlab{}.
\newblock \showarticletitle{{Living Papers: A Language Toolkit for Augmented Scholarly Communication}}. In \bibinfo{booktitle}{\emph{Proceedings of the 36th Annual ACM Symposium on User Interface Software and Technology}}. \bibinfo{publisher}{ACM}, \bibinfo{pages}{1--13}.
\newblock


\bibitem[Heimerl et~al\mbox{.}(2016)]%
        {7192685}
\bibfield{author}{\bibinfo{person}{Florian Heimerl}, \bibinfo{person}{Qi Han}, \bibinfo{person}{Steffen Koch}, {and} \bibinfo{person}{Thomas Ertl}.} \bibinfo{year}{2016}\natexlab{}.
\newblock \showarticletitle{CiteRivers: Visual Analytics of Citation Patterns}.
\newblock \bibinfo{journal}{\emph{IEEE Transactions on Visualization and Computer Graphics}} \bibinfo{volume}{22}, \bibinfo{number}{1} (\bibinfo{year}{2016}), \bibinfo{pages}{190--199}.
\newblock


\bibitem[Holton(2007)]%
        {holton2007coding}
\bibfield{author}{\bibinfo{person}{Judith~A Holton}.} \bibinfo{year}{2007}\natexlab{}.
\newblock \showarticletitle{The coding process and its challenges}.
\newblock \bibinfo{journal}{\emph{The Sage handbook of grounded theory}}  \bibinfo{volume}{3} (\bibinfo{year}{2007}), \bibinfo{pages}{265--289}.
\newblock


\bibitem[Hyland(2011)]%
        {hyland2011presentation}
\bibfield{author}{\bibinfo{person}{Ken Hyland}.} \bibinfo{year}{2011}\natexlab{}.
\newblock \showarticletitle{The presentation of self in scholarly life: Identity and marginalization in academic homepages}.
\newblock \bibinfo{journal}{\emph{English for Specific Purposes}} \bibinfo{volume}{30}, \bibinfo{number}{4} (\bibinfo{year}{2011}), \bibinfo{pages}{286--297}.
\newblock


\bibitem[Hyland(2018)]%
        {hyland2018narrative}
\bibfield{author}{\bibinfo{person}{Ken Hyland}.} \bibinfo{year}{2018}\natexlab{}.
\newblock \showarticletitle{Narrative, identity and academic storytelling}.
\newblock \bibinfo{journal}{\emph{ILCEA. Revue de l’Institut des langues et cultures d'Europe, Am{\'e}rique, Afrique, Asie et Australie}} \bibinfo{number}{31} (\bibinfo{year}{2018}).
\newblock


\bibitem[Kang et~al\mbox{.}(2022)]%
        {kang2022threddy}
\bibfield{author}{\bibinfo{person}{Hyeonsu Kang}, \bibinfo{person}{Joseph~Chee Chang}, \bibinfo{person}{Yongsung Kim}, {and} \bibinfo{person}{Aniket Kittur}.} \bibinfo{year}{2022}\natexlab{}.
\newblock \showarticletitle{Threddy: An interactive system for personalized thread-based exploration and organization of scientific literature}. In \bibinfo{booktitle}{\emph{Proceedings of the 35th Annual ACM Symposium on User Interface Software and Technology}}. \bibinfo{pages}{1--15}.
\newblock


\bibitem[Kang et~al\mbox{.}(2023)]%
        {kang2023synergi}
\bibfield{author}{\bibinfo{person}{Hyeonsu~B Kang}, \bibinfo{person}{Tongshuang Wu}, \bibinfo{person}{Joseph~Chee Chang}, {and} \bibinfo{person}{Aniket Kittur}.} \bibinfo{year}{2023}\natexlab{}.
\newblock \showarticletitle{Synergi: A mixed-initiative system for scholarly synthesis and sensemaking}. In \bibinfo{booktitle}{\emph{Proceedings of the 36th Annual ACM Symposium on User Interface Software and Technology}}. \bibinfo{pages}{1--19}.
\newblock


\bibitem[Kurosawa and Takama(2011)]%
        {kurosawa2011predicting}
\bibfield{author}{\bibinfo{person}{Takeshi Kurosawa} {and} \bibinfo{person}{Yasufumi Takama}.} \bibinfo{year}{2011}\natexlab{}.
\newblock \showarticletitle{Predicting researchers' future activities using visualization system for co-authorship networks}. In \bibinfo{booktitle}{\emph{2011 IEEE/WIC/ACM International Conferences on Web Intelligence and Intelligent Agent Technology}}, Vol.~\bibinfo{volume}{1}. IEEE, \bibinfo{pages}{332--339}.
\newblock


\bibitem[Latif and Beck(2018)]%
        {latif2018vis}
\bibfield{author}{\bibinfo{person}{Shahid Latif} {and} \bibinfo{person}{Fabian Beck}.} \bibinfo{year}{2018}\natexlab{}.
\newblock \showarticletitle{VIS Author Profiles: Interactive descriptions of publication records combining text and visualization}.
\newblock \bibinfo{journal}{\emph{IEEE transactions on visualization and computer graphics}} \bibinfo{volume}{25}, \bibinfo{number}{1} (\bibinfo{year}{2018}), \bibinfo{pages}{152--161}.
\newblock


\bibitem[Learning(2011)]%
        {learning2011apolo}
\bibfield{author}{\bibinfo{person}{Machine Learning}.} \bibinfo{year}{2011}\natexlab{}.
\newblock \showarticletitle{Apolo: Making Sense of Large Network Data by Combining}.
\newblock  (\bibinfo{year}{2011}).
\newblock


\bibitem[Lee et~al\mbox{.}(2005)]%
        {10.1145/1056808.1057069}
\bibfield{author}{\bibinfo{person}{Bongshin Lee}, \bibinfo{person}{Mary Czerwinski}, \bibinfo{person}{George Robertson}, {and} \bibinfo{person}{Benjamin~B. Bederson}.} \bibinfo{year}{2005}\natexlab{}.
\newblock \showarticletitle{Understanding research trends in conferences using paperLens}. In \bibinfo{booktitle}{\emph{CHI '05 Extended Abstracts on Human Factors in Computing Systems}} \emph{(\bibinfo{series}{CHI EA '05})}. \bibinfo{publisher}{ACM}, \bibinfo{pages}{1969–1972}.
\newblock
\showISBNx{1595930027}


\bibitem[Lee et~al\mbox{.}(2015)]%
        {Lee2015}
\bibfield{author}{\bibinfo{person}{Bongshin Lee}, \bibinfo{person}{Nathalie~Henry Riche}, \bibinfo{person}{Petra Isenberg}, {and} \bibinfo{person}{Sheelagh Carpendale}.} \bibinfo{year}{2015}\natexlab{}.
\newblock \showarticletitle{{More Than Telling a Story: Transforming Data into Visually Shared Stories}}.
\newblock \bibinfo{journal}{\emph{IEEE Computer Graphics and Applications}} \bibinfo{volume}{35}, \bibinfo{number}{5} (\bibinfo{year}{2015}), \bibinfo{pages}{84--90}.
\newblock


\bibitem[Lee et~al\mbox{.}(2024)]%
        {lee2024paperweaver}
\bibfield{author}{\bibinfo{person}{Yoonjoo Lee}, \bibinfo{person}{Hyeonsu~B Kang}, \bibinfo{person}{Matt Latzke}, \bibinfo{person}{Juho Kim}, \bibinfo{person}{Jonathan Bragg}, \bibinfo{person}{Joseph~Chee Chang}, {and} \bibinfo{person}{Pao Siangliulue}.} \bibinfo{year}{2024}\natexlab{}.
\newblock \showarticletitle{Paperweaver: Enriching topical paper alerts by contextualizing recommended papers with user-collected papers}. In \bibinfo{booktitle}{\emph{Proceedings of the 2024 CHI Conference on Human Factors in Computing Systems}}. \bibinfo{pages}{1--19}.
\newblock


\bibitem[Liu et~al\mbox{.}(2024b)]%
        {Liu2024e}
\bibfield{author}{\bibinfo{person}{Jiayu Liu}, \bibinfo{person}{Zhenya Huang}, \bibinfo{person}{Tong Xiao}, \bibinfo{person}{Jing Sha}, {and} \bibinfo{person}{Jinze Wu}.} \bibinfo{year}{2024}\natexlab{b}.
\newblock \showarticletitle{{SocraticLM: Exploring Socratic Personalized Teaching with Large Language Models}}. In \bibinfo{booktitle}{\emph{NeurIPS}}. \bibinfo{pages}{1--29}.
\newblock


\bibitem[Liu et~al\mbox{.}(2024a)]%
        {Liu2024b}
\bibfield{author}{\bibinfo{person}{Yiren Liu}, \bibinfo{person}{Si Chen}, \bibinfo{person}{Haocong Cheng}, \bibinfo{person}{Mengxia Yu}, \bibinfo{person}{Xiao Ran}, \bibinfo{person}{Andrew Mo}, \bibinfo{person}{Yiliu Tang}, {and} \bibinfo{person}{Yun Huang}.} \bibinfo{year}{2024}\natexlab{a}.
\newblock \showarticletitle{{How AI Processing Delays Foster Creativity: Exploring Research Question Co-Creation with an LLM-based Agent}}. In \bibinfo{booktitle}{\emph{Proceedings of the CHI Conference on Human Factors in Computing Systems}}. \bibinfo{publisher}{ACM}, \bibinfo{pages}{1--25}.
\newblock


\bibitem[Matejka et~al\mbox{.}(2012)]%
        {10.1145/2212776.2212796}
\bibfield{author}{\bibinfo{person}{Justin Matejka}, \bibinfo{person}{Tovi Grossman}, {and} \bibinfo{person}{George Fitzmaurice}.} \bibinfo{year}{2012}\natexlab{}.
\newblock \showarticletitle{Citeology: visualizing paper genealogy}. In \bibinfo{booktitle}{\emph{CHI '12 Extended Abstracts on Human Factors in Computing Systems}} \emph{(\bibinfo{series}{CHI EA '12})}. \bibinfo{publisher}{ACM}, \bibinfo{pages}{181–190}.
\newblock
\showISBNx{9781450310161}


\bibitem[Nguyen et~al\mbox{.}(2024)]%
        {nguyen2024human}
\bibfield{author}{\bibinfo{person}{Andy Nguyen}, \bibinfo{person}{Yvonne Hong}, \bibinfo{person}{Belle Dang}, {and} \bibinfo{person}{Xiaoshan Huang}.} \bibinfo{year}{2024}\natexlab{}.
\newblock \showarticletitle{Human-AI collaboration patterns in AI-assisted academic writing}.
\newblock \bibinfo{journal}{\emph{Studies in Higher Education}} \bibinfo{volume}{49}, \bibinfo{number}{5} (\bibinfo{year}{2024}), \bibinfo{pages}{847--864}.
\newblock


\bibitem[Ouyang et~al\mbox{.}(2024)]%
        {NotePlayer}
\bibfield{author}{\bibinfo{person}{Yang Ouyang}, \bibinfo{person}{Leixian Shen}, \bibinfo{person}{Yun Wang}, {and} \bibinfo{person}{Quan Li}.} \bibinfo{year}{2024}\natexlab{}.
\newblock \showarticletitle{{NotePlayer: Engaging Computational Notebooks for Dynamic Presentation of Analytical Processes}}. In \bibinfo{booktitle}{\emph{Proceedings of the 37th Annual ACM Symposium on User Interface Software and Technology}}. \bibinfo{publisher}{ACM}, \bibinfo{pages}{1--20}.
\newblock


\bibitem[Overney et~al\mbox{.}(2024)]%
        {overney2024sensemate}
\bibfield{author}{\bibinfo{person}{Cassandra Overney}, \bibinfo{person}{Bel{\'e}n Sald{\'\i}as}, \bibinfo{person}{Dimitra Dimitrakopoulou}, {and} \bibinfo{person}{Deb Roy}.} \bibinfo{year}{2024}\natexlab{}.
\newblock \showarticletitle{Sensemate: An accessible and beginner-friendly human-ai platform for qualitative data analysis}. In \bibinfo{booktitle}{\emph{Proceedings of the 29th International Conference on Intelligent User Interfaces}}. \bibinfo{pages}{922--939}.
\newblock


\bibitem[Palani et~al\mbox{.}(2023)]%
        {palani2023relatedly}
\bibfield{author}{\bibinfo{person}{Srishti Palani}, \bibinfo{person}{Aakanksha Naik}, \bibinfo{person}{Doug Downey}, \bibinfo{person}{Amy~X Zhang}, \bibinfo{person}{Jonathan Bragg}, {and} \bibinfo{person}{Joseph~Chee Chang}.} \bibinfo{year}{2023}\natexlab{}.
\newblock \showarticletitle{Relatedly: Scaffolding literature reviews with existing related work sections}. In \bibinfo{booktitle}{\emph{Proceedings of the 2023 CHI Conference on Human Factors in Computing Systems}}. \bibinfo{pages}{1--20}.
\newblock


\bibitem[Pang et~al\mbox{.}(2025)]%
        {10.1145/3706598.3713726}
\bibfield{author}{\bibinfo{person}{Rock~Yuren Pang}, \bibinfo{person}{Hope Schroeder}, \bibinfo{person}{Kynnedy~Simone Smith}, \bibinfo{person}{Solon Barocas}, \bibinfo{person}{Ziang Xiao}, \bibinfo{person}{Emily Tseng}, {and} \bibinfo{person}{Danielle Bragg}.} \bibinfo{year}{2025}\natexlab{}.
\newblock \showarticletitle{Understanding the LLM-ification of CHI: Unpacking the Impact of LLMs at CHI through a Systematic Literature Review}. In \bibinfo{booktitle}{\emph{Proceedings of the 2025 CHI Conference on Human Factors in Computing Systems}} \emph{(\bibinfo{series}{CHI '25})}. \bibinfo{publisher}{Association for Computing Machinery}, \bibinfo{address}{New York, NY, USA}, Article \bibinfo{articleno}{456}, \bibinfo{numpages}{20}~pages.
\newblock
\showISBNx{9798400713941}


\bibitem[Ponsard et~al\mbox{.}(2016)]%
        {ponsard2016paperquest}
\bibfield{author}{\bibinfo{person}{Antoine Ponsard}, \bibinfo{person}{Francisco Escalona}, {and} \bibinfo{person}{Tamara Munzner}.} \bibinfo{year}{2016}\natexlab{}.
\newblock \showarticletitle{PaperQuest: A visualization tool to support literature review}. In \bibinfo{booktitle}{\emph{Proceedings of the 2016 CHI Conference Extended Abstracts on Human Factors in Computing Systems}}. \bibinfo{pages}{2264--2271}.
\newblock


\bibitem[Rachatasumrit et~al\mbox{.}(2022)]%
        {rachatasumrit2022citeread}
\bibfield{author}{\bibinfo{person}{Napol Rachatasumrit}, \bibinfo{person}{Jonathan Bragg}, \bibinfo{person}{Amy~X Zhang}, {and} \bibinfo{person}{Daniel~S Weld}.} \bibinfo{year}{2022}\natexlab{}.
\newblock \showarticletitle{Citeread: Integrating localized citation contexts into scientific paper reading}. In \bibinfo{booktitle}{\emph{Proceedings of the 27th International Conference on Intelligent User Interfaces}}. \bibinfo{pages}{707--719}.
\newblock


\bibitem[Rosenthal(2006)]%
        {rosenthal2006narrated}
\bibfield{author}{\bibinfo{person}{Gabriele Rosenthal}.} \bibinfo{year}{2006}\natexlab{}.
\newblock \showarticletitle{The narrated life story: On the interrelation between experience, memory and narration}.
\newblock \bibinfo{publisher}{University of Huddersfield}.
\newblock


\bibitem[Schimel(2012)]%
        {schimel2012writing}
\bibfield{author}{\bibinfo{person}{Joshua Schimel}.} \bibinfo{year}{2012}\natexlab{}.
\newblock \bibinfo{booktitle}{\emph{Writing science: how to write papers that get cited and proposals that get funded}}.
\newblock \bibinfo{publisher}{OUP USA}.
\newblock


\bibitem[Schulze(2014)]%
        {schulze2014finding}
\bibfield{author}{\bibinfo{person}{Salom{\'e} Schulze}.} \bibinfo{year}{2014}\natexlab{}.
\newblock \showarticletitle{Finding the academic self: Identity development of academics as doctoral students}.
\newblock \bibinfo{journal}{\emph{Koers: Bulletin for Christian Scholarship= Koers: Bulletin vir Christelike Wetenskap}} \bibinfo{volume}{79}, \bibinfo{number}{1} (\bibinfo{year}{2014}), \bibinfo{pages}{1--8}.
\newblock


\bibitem[Setlur et~al\mbox{.}(2020)]%
        {Setlur2020}
\bibfield{author}{\bibinfo{person}{Vidya Setlur}, \bibinfo{person}{Enamul Hoque}, \bibinfo{person}{Dae~Hyun Kim}, {and} \bibinfo{person}{Angel~X. Chang}.} \bibinfo{year}{2020}\natexlab{}.
\newblock \showarticletitle{{Sneak pique: Exploring autocompletion as a data discovery scaffold for supporting visual analysis}}. In \bibinfo{booktitle}{\emph{Proceedings of the 33rd Annual ACM Symposium on User Interface Software and Technology, UIST'20}}. \bibinfo{publisher}{ACM}, \bibinfo{address}{Minneapolis, Minnesota, USA}, \bibinfo{pages}{966--978}.
\newblock


\bibitem[Shahaf et~al\mbox{.}(2012)]%
        {shahaf2012metro}
\bibfield{author}{\bibinfo{person}{Dafna Shahaf}, \bibinfo{person}{Carlos Guestrin}, {and} \bibinfo{person}{Eric Horvitz}.} \bibinfo{year}{2012}\natexlab{}.
\newblock \showarticletitle{Metro maps of science}. In \bibinfo{booktitle}{\emph{Proceedings of the 18th ACM SIGKDD international conference on Knowledge discovery and data mining}}. \bibinfo{pages}{1122--1130}.
\newblock


\bibitem[Shao et~al\mbox{.}(2024)]%
        {Shao2024a}
\bibfield{author}{\bibinfo{person}{Yijia Shao}, \bibinfo{person}{Yucheng Jiang}, \bibinfo{person}{Theodore~A. Kanell}, \bibinfo{person}{Peter Xu}, \bibinfo{person}{Omar Khattab}, {and} \bibinfo{person}{Monica~S. Lam}.} \bibinfo{year}{2024}\natexlab{}.
\newblock \showarticletitle{{Assisting in Writing Wikipedia-like Articles From Scratch with Large Language Models}}.
\newblock \bibinfo{journal}{\emph{Proceedings of the 2024 Conference of the North American Chapter of the Association for Computational Linguistics: Human Language Technologies, NAACL 2024}} \bibinfo{volume}{1}, \bibinfo{number}{2018} (\bibinfo{year}{2024}), \bibinfo{pages}{6252--6278}.
\newblock


\bibitem[Shao et~al\mbox{.}(2025)]%
        {NarrativePlayer}
\bibfield{author}{\bibinfo{person}{Zekai Shao}, \bibinfo{person}{Leixian Shen}, \bibinfo{person}{Haotian Li}, \bibinfo{person}{Yi Shan}, \bibinfo{person}{Huamin Qu}, \bibinfo{person}{Yun Wang}, {and} \bibinfo{person}{Siming Chen}.} \bibinfo{year}{2025}\natexlab{}.
\newblock \showarticletitle{{Narrative Player: Reviving Data Narratives with Visuals}}.
\newblock \bibinfo{journal}{\emph{IEEE Transactions on Visualization and Computer Graphics}} (\bibinfo{year}{2025}), \bibinfo{pages}{1--15}.
\newblock


\bibitem[Shen et~al\mbox{.}(2024b)]%
        {dataplaywright}
\bibfield{author}{\bibinfo{person}{Leixian Shen}, \bibinfo{person}{Haotian Li}, \bibinfo{person}{Yun Wang}, \bibinfo{person}{Tianqi Luo}, \bibinfo{person}{Yuyu Luo}, {and} \bibinfo{person}{Huamin Qu}.} \bibinfo{year}{2024}\natexlab{b}.
\newblock \showarticletitle{{Data Playwright: Authoring Data Videos With Annotated Narration}}.
\newblock \bibinfo{journal}{\emph{IEEE Transactions on Visualization and Computer Graphics}} (\bibinfo{year}{2024}), \bibinfo{pages}{1--14}.
\newblock


\bibitem[Shen et~al\mbox{.}(2024a)]%
        {datadirector}
\bibfield{author}{\bibinfo{person}{Leixian Shen}, \bibinfo{person}{Haotian Li}, \bibinfo{person}{Yun Wang}, {and} \bibinfo{person}{Huamin Qu}.} \bibinfo{year}{2024}\natexlab{a}.
\newblock \showarticletitle{{From Data to Story: Towards Automatic Animated Data Video Creation with LLM-Based Multi-Agent Systems}}. In \bibinfo{booktitle}{\emph{IEEE VIS 2024 Workshop on Data Storytelling in an Era of Generative AI, GEN4DS‘24}}. \bibinfo{publisher}{IEEE}, \bibinfo{pages}{20--27}.
\newblock


\bibitem[Shen et~al\mbox{.}(2025a)]%
        {DVSurvey}
\bibfield{author}{\bibinfo{person}{Leixian Shen}, \bibinfo{person}{Haotian Li}, \bibinfo{person}{Yun Wang}, {and} \bibinfo{person}{Huamin Qu}.} \bibinfo{year}{2025}\natexlab{a}.
\newblock \showarticletitle{{Reflecting on Design Paradigms of Animated Data Video Tools}}. In \bibinfo{booktitle}{\emph{Proceedings of the 2025 CHI Conference on Human Factors in Computing Systems}}. \bibinfo{publisher}{ACM}, \bibinfo{pages}{1--21}.
\newblock


\bibitem[Shen et~al\mbox{.}(2025b)]%
        {Instructions}
\bibfield{author}{\bibinfo{person}{Leixian Shen}, \bibinfo{person}{Haotian Li}, \bibinfo{person}{Yifang Wang}, \bibinfo{person}{Xing Xie}, {and} \bibinfo{person}{Huamin Qu}.} \bibinfo{year}{2025}\natexlab{b}.
\newblock \showarticletitle{{Prompting Generative AI with Interaction-Augmented Instructions}}. In \bibinfo{booktitle}{\emph{Extended Abstracts of the CHI Conference on Human Factors in Computing Systems, CHI EA '25}}. \bibinfo{publisher}{ACM}, \bibinfo{pages}{1--9}.
\newblock


\bibitem[Shen et~al\mbox{.}(2024c)]%
        {dataplayer}
\bibfield{author}{\bibinfo{person}{Leixian Shen}, \bibinfo{person}{Yizhi Zhang}, \bibinfo{person}{Haidong Zhang}, {and} \bibinfo{person}{Yun Wang}.} \bibinfo{year}{2024}\natexlab{c}.
\newblock \showarticletitle{{Data Player: Automatic Generation of Data Videos with Narration-Animation Interplay}}.
\newblock \bibinfo{journal}{\emph{IEEE Transactions on Visualization and Computer Graphics}} \bibinfo{volume}{30}, \bibinfo{number}{1} (\bibinfo{year}{2024}), \bibinfo{pages}{109--119}.
\newblock


\bibitem[Shi et~al\mbox{.}(2015)]%
        {7152908}
\bibfield{author}{\bibinfo{person}{Lei Shi}, \bibinfo{person}{Hanghang Tong}, \bibinfo{person}{Jie Tang}, {and} \bibinfo{person}{Chuang Lin}.} \bibinfo{year}{2015}\natexlab{}.
\newblock \showarticletitle{VEGAS: Visual influEnce GrAph Summarization on Citation Networks}.
\newblock \bibinfo{journal}{\emph{IEEE Transactions on Knowledge and Data Engineering}} \bibinfo{volume}{27}, \bibinfo{number}{12} (\bibinfo{year}{2015}), \bibinfo{pages}{3417--3431}.
\newblock


\bibitem[Stasko et~al\mbox{.}(2013)]%
        {stasko2013citevis}
\bibfield{author}{\bibinfo{person}{John Stasko}, \bibinfo{person}{Jaegul Choo}, \bibinfo{person}{Yi Han}, \bibinfo{person}{Mengdie Hu}, \bibinfo{person}{Hannah Pileggi}, \bibinfo{person}{Ramik Sadana}, {and} \bibinfo{person}{Charles~D Stolper}.} \bibinfo{year}{2013}\natexlab{}.
\newblock \showarticletitle{Citevis: Exploring conference paper citation data visually}.
\newblock \bibinfo{journal}{\emph{Posters of IEEE InfoVis}}  \bibinfo{volume}{2} (\bibinfo{year}{2013}), \bibinfo{pages}{2--3}.
\newblock


\bibitem[Strobl et~al\mbox{.}(2019)]%
        {strobl2019digital}
\bibfield{author}{\bibinfo{person}{Carola Strobl}, \bibinfo{person}{Emilie Ailhaud}, \bibinfo{person}{Kalliopi Benetos}, \bibinfo{person}{Ann Devitt}, \bibinfo{person}{Otto Kruse}, \bibinfo{person}{Antje Proske}, {and} \bibinfo{person}{Christian Rapp}.} \bibinfo{year}{2019}\natexlab{}.
\newblock \showarticletitle{Digital support for academic writing: A review of technologies and pedagogies}.
\newblock \bibinfo{journal}{\emph{Computers \& education}}  \bibinfo{volume}{131} (\bibinfo{year}{2019}), \bibinfo{pages}{33--48}.
\newblock


\bibitem[Subramonyam et~al\mbox{.}(2024)]%
        {10.1145/3613904.3642754}
\bibfield{author}{\bibinfo{person}{Hari Subramonyam}, \bibinfo{person}{Roy Pea}, \bibinfo{person}{Christopher Pondoc}, \bibinfo{person}{Maneesh Agrawala}, {and} \bibinfo{person}{Colleen Seifert}.} \bibinfo{year}{2024}\natexlab{}.
\newblock \showarticletitle{Bridging the Gulf of Envisioning: Cognitive Challenges in Prompt Based Interactions with LLMs}. In \bibinfo{booktitle}{\emph{Proceedings of the 2024 CHI Conference on Human Factors in Computing Systems}} (Honolulu, HI, USA) \emph{(\bibinfo{series}{CHI '24})}. \bibinfo{publisher}{Association for Computing Machinery}, \bibinfo{address}{New York, NY, USA}, Article \bibinfo{articleno}{1039}, \bibinfo{numpages}{19}~pages.
\newblock
\showISBNx{9798400703300}


\bibitem[Suh et~al\mbox{.}(2024a)]%
        {suh2024luminate}
\bibfield{author}{\bibinfo{person}{Sangho Suh}, \bibinfo{person}{Meng Chen}, \bibinfo{person}{Bryan Min}, \bibinfo{person}{Toby Jia-Jun Li}, {and} \bibinfo{person}{Haijun Xia}.} \bibinfo{year}{2024}\natexlab{a}.
\newblock \showarticletitle{Luminate: Structured generation and exploration of design space with large language models for human-ai co-creation}. In \bibinfo{booktitle}{\emph{Proceedings of the 2024 CHI Conference on Human Factors in Computing Systems}}. \bibinfo{pages}{1--26}.
\newblock


\bibitem[Suh et~al\mbox{.}(2024b)]%
        {10.1145/3672539.3686706}
\bibfield{author}{\bibinfo{person}{Sangho Suh}, \bibinfo{person}{Hai Dang}, \bibinfo{person}{Ryan Yen}, \bibinfo{person}{Josh~M. Pollock}, \bibinfo{person}{Ian Arawjo}, \bibinfo{person}{Rubaiat~Habib Kazi}, \bibinfo{person}{Hariharan Subramonyam}, \bibinfo{person}{Jingyi Li}, \bibinfo{person}{Nazmus Saquib}, {and} \bibinfo{person}{Arvind Satyanarayan}.} \bibinfo{year}{2024}\natexlab{b}.
\newblock \showarticletitle{Dynamic Abstractions: Building the Next Generation of Cognitive Tools and Interfaces}. In \bibinfo{booktitle}{\emph{Adjunct Proceedings of the 37th Annual ACM Symposium on User Interface Software and Technology}} (Pittsburgh, PA, USA) \emph{(\bibinfo{series}{UIST Adjunct '24})}. \bibinfo{publisher}{Association for Computing Machinery}, \bibinfo{address}{New York, NY, USA}, Article \bibinfo{articleno}{91}, \bibinfo{numpages}{3}~pages.
\newblock
\showISBNx{9798400707186}


\bibitem[Sultanum et~al\mbox{.}(2020)]%
        {sultanum2020understanding}
\bibfield{author}{\bibinfo{person}{Nicole Sultanum}, \bibinfo{person}{Christine Murad}, {and} \bibinfo{person}{Daniel Wigdor}.} \bibinfo{year}{2020}\natexlab{}.
\newblock \showarticletitle{Understanding and supporting academic literature review workflows with litsense}. In \bibinfo{booktitle}{\emph{Proceedings of the 2020 International Conference on Advanced Visual Interfaces}}. \bibinfo{pages}{1--5}.
\newblock


\bibitem[Thimbleby(2004)]%
        {thimbleby2004supporting}
\bibfield{author}{\bibinfo{person}{Harold Thimbleby}.} \bibinfo{year}{2004}\natexlab{}.
\newblock \showarticletitle{Supporting diverse HCI research}. In \bibinfo{booktitle}{\emph{Proceedings BCS HCI Conference}}, Vol.~\bibinfo{volume}{2}. \bibinfo{pages}{125--128}.
\newblock


\bibitem[Van~Eck and Waltman(2014)]%
        {van2014citnetexplorer}
\bibfield{author}{\bibinfo{person}{Nees~Jan Van~Eck} {and} \bibinfo{person}{Ludo Waltman}.} \bibinfo{year}{2014}\natexlab{}.
\newblock \showarticletitle{CitNetExplorer: A new software tool for analyzing and visualizing citation networks}.
\newblock \bibinfo{journal}{\emph{Journal of informetrics}} \bibinfo{volume}{8}, \bibinfo{number}{4} (\bibinfo{year}{2014}), \bibinfo{pages}{802--823}.
\newblock


\bibitem[Wang et~al\mbox{.}(2024a)]%
        {wang2024scidasynth}
\bibfield{author}{\bibinfo{person}{Xingbo Wang}, \bibinfo{person}{Samantha~L Huey}, \bibinfo{person}{Rui Sheng}, \bibinfo{person}{Saurabh Mehta}, {and} \bibinfo{person}{Fei Wang}.} \bibinfo{year}{2024}\natexlab{a}.
\newblock \showarticletitle{SciDaSynth: Interactive Structured Knowledge Extraction and Synthesis from Scientific Literature with Large Language Model}.
\newblock \bibinfo{journal}{\emph{arXiv preprint arXiv:2404.13765}} (\bibinfo{year}{2024}).
\newblock


\bibitem[Wang et~al\mbox{.}(2023)]%
        {vistalk}
\bibfield{author}{\bibinfo{person}{Yun Wang}, \bibinfo{person}{Zhitao Hou}, \bibinfo{person}{Leixian Shen}, \bibinfo{person}{Tongshuang Wu}, \bibinfo{person}{Jiaqi Wang}, \bibinfo{person}{He Huang}, \bibinfo{person}{Haidong Zhang}, {and} \bibinfo{person}{Dongmei Zhang}.} \bibinfo{year}{2023}\natexlab{}.
\newblock \showarticletitle{{Towards Natural Language-Based Visualization Authoring}}.
\newblock \bibinfo{journal}{\emph{IEEE Transactions on Visualization and Computer Graphics}} \bibinfo{volume}{29}, \bibinfo{number}{1} (\bibinfo{year}{2023}), \bibinfo{pages}{1222 -- 1232}.
\newblock


\bibitem[Wang et~al\mbox{.}(2024b)]%
        {wonderflow}
\bibfield{author}{\bibinfo{person}{Yun Wang}, \bibinfo{person}{Leixian Shen}, \bibinfo{person}{Zhengxin You}, \bibinfo{person}{Xinhuan Shu}, \bibinfo{person}{Bongshin Lee}, \bibinfo{person}{John Thompson}, \bibinfo{person}{Haidong Zhang}, {and} \bibinfo{person}{Dongmei Zhang}.} \bibinfo{year}{2024}\natexlab{b}.
\newblock \showarticletitle{{WonderFlow: Narration-Centric Design of Animated Data Videos}}.
\newblock \bibinfo{journal}{\emph{IEEE Transactions on Visualization and Computer Graphics}} (\bibinfo{year}{2024}), \bibinfo{pages}{1--17}.
\newblock
\showISSN{1077-2626}


\bibitem[Wang et~al\mbox{.}(2018)]%
        {wang2018visualizing}
\bibfield{author}{\bibinfo{person}{Yong Wang}, \bibinfo{person}{Conglei Shi}, \bibinfo{person}{Liangyue Li}, \bibinfo{person}{Hanghang Tong}, {and} \bibinfo{person}{Huamin Qu}.} \bibinfo{year}{2018}\natexlab{}.
\newblock \showarticletitle{{Visualizing Research Impact through Citation Data}}.
\newblock \bibinfo{journal}{\emph{ACM Transactions on Interactive Intelligent Systems}} \bibinfo{volume}{8}, \bibinfo{number}{1} (\bibinfo{year}{2018}), \bibinfo{pages}{1--24}.
\newblock
\showISSN{2160-6455}


\bibitem[Wu et~al\mbox{.}(2013)]%
        {10.1145/2492517.2492638}
\bibfield{author}{\bibinfo{person}{Meng Qi~Yelena Wu}, \bibinfo{person}{Robert Faris}, {and} \bibinfo{person}{Kwan-Liu Ma}.} \bibinfo{year}{2013}\natexlab{}.
\newblock \showarticletitle{Visual exploration of academic career paths}. In \bibinfo{booktitle}{\emph{Proceedings of the 2013 IEEE/ACM International Conference on Advances in Social Networks Analysis and Mining}} \emph{(\bibinfo{series}{ASONAM '13})}. \bibinfo{publisher}{ACM}, \bibinfo{pages}{779–786}.
\newblock


\bibitem[Xiao(2023)]%
        {Xiao2023a}
\bibfield{author}{\bibinfo{person}{Chang Xiao}.} \bibinfo{year}{2023}\natexlab{}.
\newblock \showarticletitle{{AutoSurveyGPT: GPT-Enhanced Automated Literature Discovery}}. In \bibinfo{booktitle}{\emph{Adjunct Proceedings of the 36th Annual ACM Symposium on User Interface Software and Technology}}. \bibinfo{publisher}{ACM}, \bibinfo{pages}{1--3}.
\newblock


\bibitem[Yao et~al\mbox{.}(2007)]%
        {yao2007unified}
\bibfield{author}{\bibinfo{person}{Limin Yao}, \bibinfo{person}{Jie Tang}, {and} \bibinfo{person}{Juanzi Li}.} \bibinfo{year}{2007}\natexlab{}.
\newblock \showarticletitle{A unified approach to researcher profiling}. In \bibinfo{booktitle}{\emph{IEEE/WIC/ACM International Conference on Web Intelligence (WI'07)}}. IEEE, \bibinfo{pages}{359--366}.
\newblock


\bibitem[Zhao et~al\mbox{.}(2013)]%
        {6634163}
\bibfield{author}{\bibinfo{person}{Jian Zhao}, \bibinfo{person}{Christopher Collins}, \bibinfo{person}{Fanny Chevalier}, {and} \bibinfo{person}{Ravin Balakrishnan}.} \bibinfo{year}{2013}\natexlab{}.
\newblock \showarticletitle{Interactive Exploration of Implicit and Explicit Relations in Faceted Datasets}.
\newblock \bibinfo{journal}{\emph{IEEE Transactions on Visualization and Computer Graphics}} \bibinfo{volume}{19}, \bibinfo{number}{12} (\bibinfo{year}{2013}), \bibinfo{pages}{2080--2089}.
\newblock


\bibitem[Zheng et~al\mbox{.}(2024)]%
        {zheng2024disciplink}
\bibfield{author}{\bibinfo{person}{Chengbo Zheng}, \bibinfo{person}{Yuanhao Zhang}, \bibinfo{person}{Zeyu Huang}, \bibinfo{person}{Chuhan Shi}, \bibinfo{person}{Minrui Xu}, {and} \bibinfo{person}{Xiaojuan Ma}.} \bibinfo{year}{2024}\natexlab{}.
\newblock \showarticletitle{DiscipLink: Unfolding Interdisciplinary Information Seeking Process via Human-AI Co-Exploration}. In \bibinfo{booktitle}{\emph{Proceedings of the 37th Annual ACM Symposium on User Interface Software and Technology}}. \bibinfo{pages}{1--20}.
\newblock


\end{thebibliography}
%\clearpage

\appendix
\onecolumn
\newpage
\section{Prompt for top-down exploration (linear example)}
\label{appendix1}
%In the top-down workflow, the prompt chain is designed to guide the language model through structured reasoning, informed by narrative frameworks. It includes three main modules: 1) a structure-specific instruction module, 2) a stepwise reasoning module, and 3) a schema-conforming output module. 

\begin{lstlisting}

You are an AI assistant and an expert in Human Computer Interaction research. Your task is to help researchers structure a set of papers and craft coherent research narratives.

# DATA & CONTEXT  
- Paper set               : {paper_set}  
- Overall research focus  : {overall_focus}  
- Researcher's intent     : {researcher_intent}  

# KEY CONCEPTS  
- Contribution Statement: a high-level criterion that partitions the entire paper set into meaningful, non-overlapping dimensions.  
- Cluster: a subgroup of papers that share a distinctive feature under the same Contribution Statement.  
- Cluster Theme: the abstract feature (e.g., method, technology, user group) that defines a Cluster.  
- Paper: an individual publication, referenced only by its ID.  

## Hierarchy
1. Contribution Statement > Clusters > Papers. Each Paper appears in every Contribution Statement but in exactly one Cluster per Contribution Statement (no overlaps). Together, the Clusters must cover the entire paper set.  

## Rules for Linear Framework  
Within one Contribution Statement, Clusters illustrate a sequential flow where papers are organized in a progressive line, highlighting how each stage builds upon insights from previous stages. Therefore, the Linear Framework is ideal for showcasing how initial explorations led to sophisticated outcomes or how works span different levels of advancement.

# INSTRUCTIONS  
## Step 1: Differentiate Features Among Papers: 
Examine each Paper to identify salient features (topics, technologies, research methodologies, target groups, interaction techniques, etc.).  

## Step 2: Form Preliminary Clusters and Abstract Cluster Themes
Group similar Papers into 3 to 6 Clusters. For each Cluster provide: 1) a concise, descriptive name (cluster_theme), and 2) a list of Paper IDs

## Step 3: Synthesize Contribution Statements  
Review the preliminary Clusters and abstract 4 to 6 higher-level, fundamental criteria that distinguish them. 
These criteria become the contribution_statements.  

## Important Constraints  
- Focus on distinctions that genuinely help researchers articulate unique contributions.  
- Avoid excessively broad, vague, or redundant categories.  

# OUTPUT FORMAT (strict JSON)  
Return only valid JSON. No extra keys, comments, or text. Follow exactly the schema below. Use only Paper IDs inside "papers_assign".  

{
  "contribution_statements": [
    {
      "contribution_statement": "Contribution Statement 1",
      "contribution_statement_description": "One-sentence explanation of this Contribution Statement and how its Clusters form a progression.",
      "clusters": [
        {
          "cluster_theme": "Cluster 1 Theme",
          "cluster_description": "Brief explanation of the Cluster Theme.",
          "papers_assign": ["id", "id"]
        },
        {
          "cluster_theme": "Cluster 2 Theme",
          "cluster_description": "Brief explanation of the Cluster Theme.",
          "papers_assign": ["id", "id"]
        }
      ]
    },
    {
      "contribution_statement": "Contribution Statement 2",
      "contribution_statement_description": "...",
      "clusters": [
        {
          "cluster_theme": "Cluster 1 Theme",
          "cluster_description": "...",
          "papers_assign": ["id"]
        }
      ]
    }
  ]
}

\end{lstlisting}

\section{Prompt for bottom-up exploration (linear example)}
\label{appendix2}
\begin{lstlisting}

You are an AI assistant and an expert in Human-Computer Interaction research. You are given a structured research narrative that organizes papers into contribution statements and clusters. Your task is to refine the specific field indicated by key_to_modify, using the surrounding context for reasoning.


# DATA & CONTEXT
The input is a JSON snippet that belongs to the following hierarchy: Contribution Statement > Clusters > Papers

# KEYS USED
- contribution_statement: A high-level category that groups multiple research themes  
- cluster_themeX: A specific research theme that falls under the contribution_statement  
- papers_assignX: A list of paper IDs assigned to the corresponding cluster  
- key_to_modify: The key whose value should be updated

# INPUT FORMAT
{
  "contribution_statement": "{contribution_statement}",
  "cluster_theme0": "{cluster_theme0}",
  "papers_assign0": ["id", "id"],
  "cluster_theme1": "{cluster_theme1}",
  "papers_assign1": ["id"],
  "...": "...",
  "key_to_modify": "{key_to_modify}"
}

# INSTRUCTIONS
## Step 1: Examine the Context of key_to_modify: 
- If it is a cluster_themeX: Examine the papers in the associated papers_assignX. Identify their shared characteristics (such as research method, user group, interaction type, or technology), and summarize them into a concise, meaningful cluster theme that reflects the essence of this group.
- If it is a papers_assignX: Based on the corresponding cluster_themeX and the overall contribution_statement, assign the most relevant papers from the dataset to this cluster. Ensure that the assignment is coherent with the theme and mutually exclusive with other clusters under the same contribution_statement.  
- If it is a contribution_statement: Consider all available cluster themes and papers to propose a more coherent and meaningful statement.

## Step 2: Check the Relationship Among Clusters:
- Check that the Clusters illustrate a sequential flow where papers are organized in a progressive line, highlighting how each stage builds upon insights from previous stages. Make sure the Clusters are ideal for showcasing how initial explorations led to sophisticated outcomes or how works span different levels of advancement. 
- Confirm that each paper appears in exactly one cluster within this contribution_statement.

## Step 3: Update the New Value
Locate the field indicated by key_to_modify. Replace its value with an improved one-line description. Do not modify any other fields.

# OUTPUT
Return only the new value you assign to the key. No extra keys, comments, or text.

{new_value}

\end{lstlisting}

\end{document}